\documentclass[a4paper,11pt]{article}
\usepackage{articlestyle}
\usepackage[scale={0.8,0.9},centering,includeheadfoot]{geometry}
\usepackage{paralist}

\usepackage{articlestyle}

\graphicspath{{figs/}}
\title{Excursion and contour uncertainty regions\\ for latent Gaussian models}
\date{}
\begin{document}

\begin{center}
\textbf{\textsf{\huge
Excursion and contour uncertainty regions\\ for latent Gaussian models}}\\
\vspace{5mm}
{\Large \scshape{David Bolin$^1$ and Finn Lindgren$^2$}}

\vspace{3mm}

\textit{$^1$\matcent} \\
\textit{$^2$\bath}

\vspace{3mm}

\begin{minipage}{0.9\textwidth}
{\small
{\bf Abstract:} An interesting statistical problem is to find regions where some studied process exceeds a certain level. Estimating such regions so that the probability for exceeding the level in the entire set is equal to some predefined value is a difficult problem that occurs in several areas of applications ranging from brain imaging to astrophysics. In this work, a method for solving this problem, as well as the related problem of finding uncertainty regions for contour curves, for latent Gaussian models is proposed. The method is based on using a parametric family for the excursion sets in combination with a sequential importance sampling method for estimating joint probabilities. The accuracy of the method is investigated using simulated data and two environmental applications are presented. In the first application, areas where the air pollution in the Piemonte region in northern Italy exceeds the daily limit value, set by the European Union for human health protection, are estimated. In the second application, regions in the African Sahel that experienced an increase in vegetation after the drought period in the early 1980s are estimated.

\begin{flushleft}
{\bf Key words:}
Latent Gaussian models; excursion sets; contour curves; multiple testing
\end{flushleft}
}
\end{minipage}
\end{center}

\section{Introduction}
In many statistical applications, one is interested in finding areas where the studied process exceeds a certain level or is significantly different from some reference level. A typical example is in studies of air pollution, where one is interested in testing if, and where, the pollution level exceeds some given limit value set by some regulatory agency \citep{cameletti12}, and similar examples can be found in a wide range of scientific fields including brain imaging \citep{marchini03} and astrophysics \citep{beaky92}. In spatio-temporal applications one might be interested in finding regions that have experienced significant changes over the studied time period. This is a common problem in climate science and the studied quantity can for example be temperature \citep{furrer07}, precipitation \citep{sain11}, or vegetation \citep{eklundh,bolin09a}.

The quintessential problem is that one has observations $\mv{y}$ of some latent stochastic field $x(\mv{s})$ and wants to find a region $D$ such that, with a certain given probability, $x(\mv{s})>u$ for all $\mv{s}\in D$ for a given level $u$. The easiest, and most common, way of specifying $D$ is to choose it as the set of locations
\begin{equation}\label{eq:marginalset}
D_m = \{\mv{s}: \pP(x(\mv{s})>u)\geq 1-\alpha\},
\end{equation}
where the probability is taken under the posterior distribution for $x|\mv{y}$.
Thus, $D_m$ is specified as the set of locations where the marginal probability for exceeding the level exceeds some given value $1-\alpha$. The set can be calculated using pointwise hypothesis testing and the parameter $\alpha$ then acts as the type 1 error parameter, and thus controls how confident one is that the level is exceeded at each location. The problem with this definition of $D$ is that of multiple hypothesis testing; the confidence level $\alpha$ does not give us any information about the family-wise error rate, and hence does not quantify the certainty of the level being exceeded at all points in the set simultaneously. That is, the probability $\pP(x(\mv{s})>u, \mv{s}\in D_m)$ is in general smaller than $1-\alpha$. If one instead wants to choose $D$ so that this simultaneous probability is $1-\alpha$, one has to modify the procedure for constructing the set. 

The more general problem of multiple hypothesis testing is an active research area and there exists a number of proposed solutions for problems in various contexts. Most of these solutions are based on first calculating the marginal probabilities $\pP(x(\mv{s})>u)$, then calculating a single threshold value $t$, and finally defining the exceedance region as $D = \{\mv{s}: \pP(x(\mv{s})>u)>t\}$. The methods differ in how the threshold $t$ is calculated, and can basically be divided into three main categories; type 1 error control thresholding, false discovery rate thresholding, and posterior probability thresholding \citep{marchini03}. The most popular method is likely the method by \cite{adler81} using the Euler characteristic of the latent field to control the family-wise error rate when defining the threshold~$t$. Though this method is simple to use, one has to be careful to check whether the required assumptions are satisfied. Typically the method is accurate for large values of $u$, and the latent field is assumed to be stationary.

In this work we will focus on the problem where the latent spatial field $x(\mv{s})$ is Gaussian and measured at a set of irregular locations. This means that the posterior distribution $\pi(x|\mv{y})$ is non-stationary and typically non-Gaussian unless the measurements are Gaussian and the model parameters are known a priori. One of our motivating examples is the problem of finding regions with significant changes in vegetation in the African Sahel studied by \cite{eklundh} and \cite{bolin09a}, and in this example the threshold $u$ is zero, so methods based on asymptotic arguments when $u$ goes to infinity are unlikely to perform well. 

The method derived here is based on using a parametric family for the excursion sets in combination with a sequential importance sampling method for estimating joint probabilities. For a specific choice of the parametric family, the method is equivalent to the thresholding methods mentioned above, with the important difference that the correct joint distribution is used when selecting the threshold. The method is extended using more general parametric families, and the related problem of finding uncertainty regions for contour curves is treated using the same methodology. 

The structure of the article is as follows. In Section \ref{sec:problem}, 
the problem is formulated and definitions for excursion sets and uncertainty regions for contour curves are given. In Section~\ref{sec:computations}, a method for estimating these sets is proposed. Estimating the sets is the most difficult problem as one easily runs into computational difficulties arising from having to evaluate high-dimensional integrals. In Section~\ref{sec:simulations}, the methods are tested on a few simulated examples to test the method's accuracy. Two applications to real data are covered in Section~\ref{sec:applications}, the first considers air pollution data from the North-Italian region Piemonte, and the second considers estimation of spatially dependent vegetation trends in the African Sahel. Finally, a few remarks and comments are given in Section~\ref{sec:conclusions}.

\section{Problem formulation}\label{sec:problem}
There are a number of different ways one could formulate excursion sets, and not all of them are useful from a practical point of view. Hence, in this section we will formalise the problem and discuss how the results should be interpreted. More precisely we look at two connected problems. The first one is to find areas where a stochastic process exceeds a given level with some probability and the second one is to quantify the uncertainty in contour curves of stochastic fields. 

Throughout this section, let $\Omega$ be a bounded domain of $\R^n$, or have a well-defined area $|\Omega|<\infty$. First some notation for excursion sets of a function and contour sets is needed. 

\begin{defn}[Excursion sets for functions]
Given a function $f(\mv{s})$, $\mv{s}\in\Omega$, the positive excursion set
$A_u^+$ for a level $u$ is given by
  \begin{align*}
    A_u^+(f) &= \{ \mv{s}\in\Omega; f(\mv{s})>u \} .
  \end{align*}
Similarly
  \begin{align*}
    A_u^-(f) &= \{ \mv{s}\in\Omega; f(\mv{s})<u \} .
  \end{align*}
is the negative excursion set.
\end{defn}
In a similar fashion one could now define the set of contour points for the level $u$ as the set of points $\mv{s}$ for which $f(\mv{s})=u$; however, a contour curve consists not only of these points but also discontinuous crossings of the level $u$. In order to incorporate both continuous and discontinuous crossings, a contour point is defined as a point $\mv{s}$ such that in every neighborhood $B$ of $\mv{s}$
\begin{equation*}
\exists\, \mv{s}_1, \mv{s}_2 \in B: f(\mv{s}_1)\geq u, f(\mv{s}_2)\leq u.
\end{equation*}
The set of all such points is the complement of the union of the interior sets of the positive and negative excursion sets.
\begin{defn}[Contour sets for functions]
Given a function $f(\mv{s})$, $\mv{s}\in\Omega$, the contour set $A_u^o$ for a level $u$ is given by
  \begin{align*}
    A_u^c(f) &= \left(A_u^+(f)^o\cup A_u^-(f)^o\right)^c.
  \end{align*}
  where $A^o$ is the interior, relative to $\Omega$, of the set $A$ and $A^c$ is the complement.
\end{defn}
\begin{rem}
Taking the interiors of the sets $A_u^+(f)$ and $A_u^-(f)$ is important. Consider for example the following function on $\Omega = [0,1]$ 
\begin{equation*}
f(s) = \begin{cases}
-1 & 0 \leq s < 0.5 \\
1 & 0.5 \leq s \leq 1.
\end{cases}
\end{equation*}
In this case $A_0^+(f)\cup A_0^-(f) = \Omega$, so without taking the interiors of the sets $A_0^c(f)$ would be empty when we want to include the discontinuous crossing at $0.5$ in the contour set. It is also important to take the interiors with respect to $\Omega$ and not $\R$, since the endpoints $0$ and $1$ always would be included in the contour set otherwise. This may seem as only a theoretical nicety, but the problem with discontinuous functions occurs frequently in environmental applications when discontinuous covariates are used for the mean value function of the field.  This makes it essential to not treat contour sets as regions where the function lies close to a level, but rather as regions where level \emph{crossings} occur.
\end{rem}

The statistical problem is now to find a region $D$ such that the function $x(\mv{s})$ exceeds the level $u$ with a certain probability $1-\alpha$ for all $\mv{s}\in D$. There might be many such regions, so if one is interested in a single answer one might look for the largest of these.

\begin{defn}[Excursion sets]
Let $x(\mv{s})$, $\mv{s}\in\Omega$ be a random field (or process). The positive level $u$ excursion set with probability $1-\alpha$ is given by
  \begin{equation*}
\exset{u,\alpha}{+}(x) = \argmax_{D}\{|D| : \pP(D \subseteq A_u^+(x)) \geq 1-\alpha\}.
  \end{equation*}
  Similarly
  \begin{equation*}
\exset{u,\alpha}{-}(x) = \argmax_{D}\{|D| : \pP(D \subseteq A_u^-(x)) \geq 1-\alpha\}.
  \end{equation*}  
  is the negative level $u$ excursion set with probability $1-\alpha$.
\end{defn}
\begin{rem}
The set $\exset{u,\alpha}{+}(x)$ can also be formulated as the largest set $D$ for which $\pP(\inf_{\mv{s}\in D} x(\mv{s}) \leq u) \leq \alpha$, which can be useful when calculating the set in practice. Also note that for deterministic functions $f$ one has $\exset{u,\alpha}{+}(f) = A^+_{u}(f)$ and $\exset{u,\alpha}{-}(f) = A^-_{u}(f)$ for any $\alpha \in [0,1]$.
\end{rem}

It is important to realize how the excursion set $\exset{u,\alpha}{+}(x)$ should be interpreted: It is the largest set so that the level $u$ is exceeded \emph{at all locations} in the set with probability $1-\alpha$, and therefore is a smaller set than $D_m$ defined in \eqref{eq:marginalset}, which is the set of points where the marginal probability for exceeding the level is at least $1-\alpha$. Another possible definition of an excursion set would be a set that contains \emph{all excursions} with probability $1-\alpha$. This is a larger set than $D_m$, given by $\exset{u,\alpha}{-}(x)^c$. Which set one is interested in depends on the application, but it can be a good idea to calculate both to get a better understanding of the uncertainties in the problem. 

In certain applications, one might be interested in joint positive and negative excursions from some level, for example when doing simultaneous regressions and one is interested in finding regions where the slopes are significantly different from zero (see Section \ref{paperE:sec:sahel} for a possible scenario of this kind).

\begin{defn}[Level avoiding sets]\label{def:avoiding}
Let $x(\mv{s})$, $\mv{s}\in\Omega$ be a random field. The pair of level $u$ avoiding sets with probability $1-\alpha$ is given by
  \begin{equation*}
(M_{u,\alpha}^+(x), M_{u,\alpha}^-(x)) = \argmax_{(D^+,D^-)}\{|D^-\cup D^+| :
\pP(D^- \subseteq A_u^-(x),\, D^+ \subseteq A_u^+(x))
\geq 1-\alpha\}.
  \end{equation*}
Denote the union of these two sets the level avoiding set $M_{u,\alpha}$:
\begin{equation*}
M_{u,\alpha}(x) = M_{u,\alpha}^+(x) \cup M_{u,\alpha}^-(x).
\end{equation*}
\end{defn}
\begin{rem}
The sets $M_{u,\alpha}^+(x)$ and $M_{u,\alpha}^-(x)$ must be non-overlapping for the probability to be non-zero. The set $M_{u,\alpha}$ can be calculated as an excursion set itself. To see this, define a new random process $y(\mv{s})$ by
\begin{align*}
y(\mv{s}) &=
\begin{cases}
u-x(\mv{s}), & \mv{s}\in D^-, \\
x(\mv{s})-u, & \mv{s}\not\in D^-.
\end{cases}
\end{align*}
The probability calculation in Definition \ref{def:avoiding} can now be reformulated as an ordinary excursion probability in $y$:
\begin{align*}
\pP(D^- \subseteq A_u^-(x),\, D^+ \subseteq A_u^+(x)) &=
\pP(D^- \cup D^+ \subseteq A_0^+(y)).
\end{align*}
Also in this case, the set can be found using a reformulation using the infimum over the region as $\pP(\inf_{\mv{s}\in D^-\cup D^+} y(\mv{s}) \leq 0) \leq \alpha$.
\end{rem}

Similarly to how the contour sets for deterministic functions were defined, the pair of level avoiding sets can now be used to define uncertainty regions for contour curves.
\begin{defn}[Uncertainty region for contour sets] 
Let $x(\mv{s})$, $\mv{s}\in\Omega$ be a random field, and let $(M_{u,\alpha}^+(x), M_{u,\alpha}^-(x))$ be the pair of level avoiding sets from Definition \ref{def:avoiding}. The set
  \begin{align*}
M^c_{u,\alpha}(x) &= \left(M_{u,\alpha}^+(x)^o\cup M_{u,\alpha}^-(x)^o\right)^c
  \end{align*}
is then an uncertainty region for the contour set of level $u$.
\end{defn}
The interpretation of this uncertainty region is important. The set $M^c_{u,\alpha}$ is the smallest set such that \emph{all} level $u$ crossings of $x$ are in the set with probability $1-\alpha$. One should note that this definition of the uncertainty region for level curves is different from some other definitions in the literature. For example, \cite{Lindgren95} define uncertainty regions as a union of intervals where each interval contains a single level crossing with probability $1-\alpha$.

It is somewhat unsatisfactory that the sets defined here are made unique by finding the largest set satisfying a certain restriction. The set $\exset{u,\alpha}{+}(x)$ is for example defined as the largest set $D$ satisfying $\pP(D \subseteq A_u^+(x)) \geq 1-\alpha$, but there are also many other smaller sets satisfying the requirement, and these are not seen if only $\exset{u,\alpha}{+}(x)$ is reported. Also, if one wants to know where the field likely exceeds the level $u$, the set $\exset{u,\alpha}{+}(x)$ might not be sufficient since it does not provide any information about the locations not contained in the set. Therefore, it would be good to have something similar to $p$-values, i.e.~the marginal probabilities of exceeding the level, but which can be interpreted simultaneously. To that end we introduce the excursion function, level avoidance function, and contour function as visual tools for answering such questions. 

\begin{defn}[Excursion functions] The positive and negative $u$ excursion functions are given by
  \begin{align*}
F_u^+(\mv{s}) &= \sup\{1-\alpha ; \mv{s}\in \exset{u,\alpha}{+} \},\\
F_u^-(\mv{s}) &= \sup\{1-\alpha ; \mv{s}\in \exset{u,\alpha}{-} \}.
  \end{align*}
Similarly, the level avoidance and contour functions are given by
    \begin{align*}
F_u(\mv{s}) &= \sup\{1-\alpha ; \mv{s}\in M_{u,\alpha} \}, \\
F_u^c(\mv{s}) &= 1-F_u(\mv{s}).
  \end{align*}
\end{defn}
These functions take values between zero and one, and for a fixed level $u$ and a fixed location $\mv{s}$, this value is equal to $1-\alpha$ for the smallest $\alpha$ such that the location is a member of the excursion set. Thus, the set $\exset{u,\alpha}{\bullet}$ can be retrieved as the $1-\alpha$ excursion set of the function $F_u^{\bullet}(s)$. The interpretation of the excursion function is therefore that if, for a given location $\mv{s}$, the function takes a value close to one, this indicates that this location is a member of the excursion sets for almost all values of $\alpha$, whereas if the value of the function is close to zero, the location is only a member of excursion sets with large values of $\alpha$ and it is therefore more unlikely that the process exceeds the level at that location.

\section{Computations}\label{sec:computations}

So far, no assumptions have been made regarding the distribution of $x(\mv{s})$, but to be able to calculate the excursion sets in practice we will now restrict ourselves to the class of latent Gaussian models, which is a popular model class with many practical applications \citep[see e.g.][]{rue09}. 
Thus, the following problem setup is assumed. Let $x(\mv{s})$ be a random field that can be written on the form
\begin{equation*}
x(\mv{s}) = \sum_{i=1}^k \beta_i f_i(\mv{s}) + z(\mv{s}) 
\end{equation*}
where $f_i(\mv{s})$ are fixed effects and $z(\mv{s})$ is a zero mean random field with covariance parameters $\mv{\theta}_1$. Further assume that both $z(\mv{s})$ and the parameter vector ${\mv{\beta} = (\beta_1,\ldots,\beta_k)^{\trsp}}$ are a priori Gaussian. 

Let $\mv{y}=(y_1,\ldots,y_m)^{\trsp}$ be a vector of measurements of the latent field with some distribution $\pi(\mv{y}|\mv{x}_{obs},\mv{\theta}_2)$, where $\mv{x}_{obs}$ is a vector containing the latent field evaluated at the measurement locations and $\mv{\theta}_2$ is a vector of parameters for the measurement distribution. Finally let $\mv{s}_1, \ldots, \mv{s}_n$ be the set of locations where predictions of the latent field should be calculated and let $\mv{x} = (x(\mv{s}_1),\ldots,x(\mv{s}_n))^{\trsp}$. The posterior distribution for $\mv{x}$ can then be written as 
\begin{equation}\label{eq:posterior}
\pi(\mv{x}|\mv{y}) = \int \pi(\mv{x}|\mv{y},\mv{\theta})\pi(\mv{\theta}|\mv{y})\md\mv{\theta},
\end{equation}
where $\mv{\theta} = (\mv{\theta}_1^{\trsp},\mv{\theta}_2^{\trsp})^{\trsp}$, and this is the distribution that should be used in the probability calculations when estimating the excursion sets. 

There are now, in principle, two main problems that have to be solved in order to find the excursion sets, level avoidance sets, or contour uncertainty sets:
\begin{description}
\item[integration] For excursion sets, calculate $\pP(D\subseteq A_u^+(\mv{x}))$ or ${\pP(D\subseteq A_u^-(\mv{x}))}$ for a given set $D$, or in the case of level avoidance sets or uncertainty regions for contour curves, calculate $\pP(D^-\subseteq A_u^-(\mv{x}), D^+\subseteq A_u^+(\mv{x}))$ for the pair of sets $(D^+,D^-)$.
\item[optimization] Use shape optimization to find largest region $D$ satisfying the required probability constraint.
\end{description}
Hence, given a method to solve each of the two problems, one could simply run the shape optimization algorithm and in each iteration calculate the required probability using the integration method. In theory there are no problems doing this, but in practice the integration method will be computationally demanding and it may not be feasible to use this strategy for applications involving large data sets. Therefore, we instead propose a slightly different strategy that will minimize the number of calls to the integration method by solving the problem sequentially. We first outline the strategy in the simplest possible situation, which will be used as a basis for all other more complicated strategies.

The method is based on using an increasing parametric family for the excursion sets in combination with a sequential integration routine for calculating the probabilities. The advantage with using a sequential integration routine is that if the required probability has been calculated for some set $D_1$, then the calculation for a larger set $D_2 \supset D_1$ can be based on the result for $D_1$, resulting in large computational savings. 

\begin{alg}[Calculating excursion sets using a one-parameter family]\label{alg1}
Assume that the model parameters $\mv{\theta}$ are known and that the posterior distribution $\pi(\mv{x}|\mv{y},\mv{\theta})$ is Gaussian. Further assume that $D(\rho)$ is a parametric family for the possible excursion sets, such that $D(\rho_1) \subseteq D(\rho_2)$ if $\rho_1 < \rho_2$. The following strategy is then used to calculate $\exset{u,\alpha}{+}$.
\begin{enumerate}
\item Choose a suitable (sequential) integration method for the problem.
\item Reorder the nodes to the order they will be added to the excursion set when the parameter $\rho$ is increased.
\item sequentially add nodes to the set $D$ according to the ordering given above and in each step update the probability $\pP(D\subseteq A_u^+(\mv{x}))$. Stop as soon as this probability falls below $1-\alpha$.
\item $\exset{u,\alpha}{+}$ is given by the last set $D$ for which $\pP(D\subseteq A_u^+(\mv{x}))\geq1-\alpha$.
\end{enumerate}
\end{alg}

The computational savings of this sequential strategy are large. For example, assume that we want to find the positive level $u$ excursion set $\exset{u,\alpha}{+}(x)$, and have $m$ candidates $D(\rho_1), \ldots, D(\rho_m)$ to choose from. Using the na\"{i}ve optimization method, we would then have to check whether $\pP(D(\rho_i)\subseteq A_u^+(x))>1-\alpha$ for each of these sets, and among the sets that satisfy the condition select the largest. Thus doing the probability calculation $m$ times. However, by reordering the nodes and adding them sequentially we only have to run the integration routine once. Also note that the excursion function $F_u^+(\mv{s})$ is retrieved by setting $\alpha=1$ and in each step saving the probabilities $\pP(D\subseteq A_u^+(\mv{x}))$. Thus, the entire excursion function can be retrieved by running the algorithm once.

Before extending this method to more general situations, we go into more detail on how to do the steps in Algorithm \ref{alg1} in practice. In Section \ref{sec:integration}, a few sequential integration methods are presented. In Section \ref{sec:families}, some different parametric families for the excursion sets and level avoidance sets are introduced and Algorithm \ref{alg1} is extended using two-parameter families. The problem of how to optimally reorder the nodes is also discussed in this section. Finally in Section \ref{sec:inlaprob}, three different methods are proposed for calculating excursion sets under the full posterior distribution \eqref{eq:posterior}.

\subsection{Gaussian probability calculations}\label{sec:integration}
For a Gaussian vector $\mv{x}$, the probabilities $\pP(D\subseteq A_u^+(\mv{x}))$, $\pP(D\subseteq A_u^-(\mv{x}))$, and $\pP(D^+\subseteq A_u^+(\mv{x}),D^-\subseteq A_u^-(\mv{x}))$ can all be written on the form 
\begin{equation}\label{eq:normalint}
I(\mv{a},\mv{b},\mv{\Sigma}) = \frac{1}{(2\pi)^{d/2}|\mv{\Sigma}|^{1/2}}\int_{\mv{a}\leq \mv{x} \leq \mv{b}}\exp(-\frac{1}{2}\mv{x}^{\trsp}\mv{\Sigma}^{-1}\mv{x})\md \mv{x},
\end{equation}
where $\mv{a}$ and $\mv{b}$ are vectors depending on the mean value of $\mv{x}$, the domain $D$, and on $u$. There have been considerable research efforts devoted to approximating integrals of this form in recent years, and we will in this section briefly describe a few techniques that can be used. 

The simplest way of approximating \eqref{eq:normalint} is to use Monte-Carlo (MC) integration. However, estimating the probability with any reasonable accuracy using standard MC integration is often too computationally expensive. Fortunately there are a number of variance reduction techniques that can be used to increase the efficiency. 

A key step in many numerical integration techniques is to transform the integral to make it more suitable for integration. Notably, \cite{Genz92} derived such a transformation for the Gaussian integral \eqref{eq:normalint}, though similar transformations have been suggested by other authors as well \citep[see e.g.][]{Geweke91}. Besides transforming the integral to the unit hyper cube, the transformation also achieves a separation of the variables so that the full problem can be calculated sequentially. The integral can then efficiently be evaluated using a quasi MC (QMC) method where the uniform random numbers in the ordinary MC integrator are replaced by some deterministic sequence of points chosen to reduce the probabilistic error bound of the crude MC integrator, see \cite{Genz09} for details.

A final variance reduction technique for the general integration problem is to reorder the variables before calculating the integral, as first suggested by \cite{Schervish84} and later improved by \cite{Gibson94}. These reorderings can reduce the error by an order of magnitude, as shown by \cite{Genz02}. However, the technique will not be applicable in our situation since the reordering will be determined by a parametric family for the excursion sets. 

\subsubsection{Methods for Markov random fields}
A common assumption in spatial statics and image analysis is that the latent field can be modeled, or approximated, using a Gaussian Markov random field (GMRF). See \cite{rue1} for an introduction to GMRFs, and note that GMRFs are used also for modeling in continuous space, for example using the SPDE approach by \cite{lindgren10}. One of the motivating reasons for using GMRFs is that it reduces the computational cost for parameter estimation and spatial prediction, and because of this one would also like to be able to use the Markov property in the calculation of \eqref{eq:normalint}. 

The main difference between latent GMRF models and standard Gaussian models is that the distribution is specified using the (sparse) precision matrix $\mv{Q}$ instead of the covariance matrix:
\begin{equation}\label{eq:markovint}
I(\mv{a},\mv{b},\mv{Q}) = \frac{|\mv{Q}|^{1/2}}{(2\pi)^{d/2}}\int_{\mv{a}\leq \mv{x} \leq \mv{b}}\exp(-\frac{1}{2}\mv{x}^{\trsp}\mv{Q} \mv{x})\md \mv{x},
\end{equation}
Using the QMC methods on GMRF models is difficult without first inverting the precision matrix and then ignoring the sparsity of $\mv{Q}$ in the calculations. To take advantage of the sparsity of $\mv{Q}$ one can use the fact that any GMRF can be viewed as a non-homogeneous auto-regressive process defined backwards in the indices of $\mv{x}$ \citep[see][Theorem 2.7]{rue1}, that is, if $\mv{x}$ is a GMRF with mean $\mv{\mu}$ and precision matrix $\mv{Q}$, then
\begin{equation}\label{eq:Qmargin}
x_i|x_{i+1},\ldots,x_{n} \sim \pN\left(\mu_i - \frac{1}{L_{ii}}\sum_{j=i+1}^n L_{ji}(x_j-\mu_j), L_{ii}^{-2}\right),
\end{equation}
where $L_{ij}$ are the elements of the Cholesky factor of $\mv{Q}$. The integral can thus be written as
\begin{equation*}
I(\mv{a},\mv{b},\mv{Q}) = \int_{a_1}^{b_1} \pi(x_1|\mv{x}_{2:d})\int_{a_2}^{b_2}\pi(x_2|\mv{x}_{3:d}) \cdots \int_{a_{d-1}}^{b_{d-1}}\pi(x_{d-1}|x_d)\int_{a_d}^{b_d}\pi(x_d)\md \mv{x}
\end{equation*}
where, because of the Markov structure, $x_i|\mv{x}_{i+1:d}$ only depends on the elements in $x_{\mathcal{N}_i\cap \{i+1:d\}}$, and $\mathcal{N}_i$ is the neighborhood of $i$ in the graph of the GMRF. 

If $\mv{Q}$ is a band-matrix, the integral can be efficiently calculated as a sequence of iterated one-dimensional integrals as discussed in \cite{genz86}. However, the band width of $\mv{L}$ will often be too large for this method to be efficient, and a better alternative is then of use a particle filter algorithm based on the GHK simulator \citep{Geweke91,Hajivassiliou93,keane93}. Denote the integral of the last $d-i$ components by $I_i$, 
\begin{equation*}
I_i = \int_{a_i}^{b_i} \pi(x_i|\mv{x}_{i+1:d}) \cdots \int_{a_{d-1}}^{b_{d-1}}\pi(x_{d-1}|x_d)\int_{a_d}^{b_d}\pi(x_d)\md \mv{x},
\end{equation*}
and note that the integral is the normalizing constant to the truncated density 
\begin{equation*}
f_i(\mv{x}_{i:d}) \propto 1(\mv{a}_{i:d}<\mv{x}_{i:d}<\mv{b}_{i:d})\pi(\mv{x}_{i:d}).
\end{equation*}
The integrals $I_d, \ldots, I_1$ are now estimated sequentially using importance sampling. In the first step, calculate ${I_d = \Phi(L_{ii}(b_d-\mu_d)) - \Phi(L_{ii}(a_d-\mu_d))}$, simulate $N$ samples $\{x_d^j\}_{j=1}^N$ from the truncated normal distribution $h_d(x_d) \propto 1(a_d<x_d<b_d)\pi(x_d)$, and set $w_d^j = I_d$. Next, simulate $x_{d-1}^j$ from ${h_{d-1}(x_{d-1}|x_d^j) = 1(a_{d-1}<x_{d-1}<b_{d-1})\pi(x_{d-1}|x_d^j)}$
and set $\mv{x}_{d-1:d}^j = \{x_{d-1}^j,x_d^j\}$. The integral $I_{d-1}$ is estimated as $I_{d-1} \approx \frac{1}{N}\sum_{j=1}^N w_{d-1}^j$ where $w_{d-1}^j$ are the importance weights 
\begin{equation}
w_{d-1}^j = \frac{1(\mv{a}_{d-1:d}<\mv{x}_{d-1:d}<\mv{b}_{d-1:d})\pi(\mv{x}_{d-1:d})}{h_{d-1}(\mv{x}_{d-1}|\mv{x}_d)h_{d}(\mv{x}_{d})}. 
\end{equation}
Proceed like this, simulating from the truncated conditional distributions and in each step updating the importance weights recursively through
\begin{equation*}
w_{i}^j = \left[\Phi\left(L_{ii}(b_i-\mu_i) + \sum_{j=i+1}^n L_{ji}(X_j-\mu_j)\right) -\Phi\left(L_{ii}(a_i-\mu_i) + \sum_{j=i+1}^n L_{ji}(X_j-\mu_j)\right)\right]w_{i+1}^j.
\end{equation*}
To reduce the variance of the estimator when the target probability is small, a resampling step can be performed after having calculated the weights $w_i^j$. This is a common strategy in particle filter techniques, and the sample $\{\mv{x}_{i:d}^j\}$ is then updated by selecting $N$ particles from the set, where $x_{i:d}^j$ is selected with probability $w_{i}^j/\sum_{k=1}^N w_{i}^k$. To avoid resampling too often, one can do the resampling only if some criterion is met, for example if the effective sample size is below some given threshold.

\subsection{Parametric families}\label{sec:families}
In theory one can use any shape optimization technique to find the largest region $D$. However, since evaluating the probability $\pP(x(\mv{s})>u, \mv{s}\in D)$ for a given set $D$ is computationally expensive, one would like to do as few iterations as possible in this step. As discussed previously, we will solve this by assuming a parametric form of the sets $D$. The optimization can then be reduced to a standard optimization of only a few variables instead of doing a full shape optimization procedure. The parametric families will be based on the marginal quantiles of $x(\mv{s})$,
  \begin{align*}
\pP(x(\mv{s}) \leq q_\rho(\mv{s})) &= \rho,
  \end{align*}
which are easy to calculate using only the marginal posterior distributions. The simplest one-parameter family based on the marginal quantiles is given in the following definition.
\begin{defn}[One-parameter family]\label{def:parametric0}
Let $q_\rho(\mv{s})$ be the marginal quantiles for $x(\mv{s})$, then a one-parameter family for the positive and negative $u$ excursion sets is given by 
  \begin{align*}
\makebox[3mm][r]{$D_1^+(\rho)$} &=
\{ \mv{s}; \pP(x(\mv{s})>u) \geq 1-\rho \} = 
\{ \mv{s}; \pP(x(\mv{s})\leq u) \leq \rho \} = 
A_u^+(q_\rho),
\\
\makebox[3mm][r]{$D_1^-(\rho)$} &=
\{ \mv{s}; \pP(x(\mv{s})<u) \geq 1-\rho \} = 
\{ \mv{s}; \pP(x(\mv{s})\geq u) \leq \rho \} = 
A_u^-(q_{1-\rho}).
  \end{align*}
\end{defn}
Using this one-parameter family in Algorithm \ref{alg1} is equivalent to finding a threshold value for the marginal excursion probabilities to get the correct simultaneous significance level. It is thus similar to the thresholding algorithms discussed in \cite{marchini03} but with the important difference that the correct joint, often non-stationary, posterior density is used when finding the threshold. 

The simple one-parameter family can be extended in a number of ways, for example by considering other levels in the excursion sets.

\begin{defn}[Two-parameter family]\label{def:parametric1}
Let $q_\rho(\mv{s})$ be the marginal quantiles for $x(\mv{s})$, then a two-parameter family for the positive and negative $u$ excursion sets is given by 
  \begin{align*}
\makebox[4.5mm][r]{$D_1^+(v,\rho)$} &=
\{ \mv{s}; \pP(x(\mv{s})>v) \geq 1-\rho \} = 
\{ \mv{s}; \pP(x(\mv{s})\leq v) \leq \rho \} = 
A_v^+(q_\rho),
\\
\makebox[4.5mm][r]{$D_1^-(v,\rho)$} &=
\{ \mv{s}; \pP(x(\mv{s})<v) \geq 1-\rho \} = 
\{ \mv{s}; \pP(x(\mv{s})\geq v) \leq \rho \} = 
A_v^-(q_{1-\rho}).
  \end{align*}
The sets $D_1^+(v,\rho)$ and $D_1^-(v,\rho)$ are increasing in $\rho$ for a fixed $v$.
\end{defn}
One drawback with this parametric family is that it does not take the spatial dependency of the data into account directly. Therefore certain sets which might seem reasonable to test are not included in the family. Consider the example in Section~\ref{sec:example1}, Figure~\ref{fig:ex1}, Panel~(a), where the marginal excursion probabilities are shown in grey for an example in one dimension where the model is a Gaussian process with exponential covariance function. The estimated posterior mean in this example is shown as the black curve in Panel (b) in the figure, and in this situation a reasonable candidate for the $0$-excursion set might be a contiguous set centered at 1, $[1-\lambda_1, 1+\lambda_2]$ for some positive $\lambda_1,\lambda_2$. However, looking at the marginal probabilities we see that sets on this form will not be included in the parametric family. One way of including such sets is to first smooth the marginal excursion probabilities $p_i=\pP(x(\mv{s}_i)>u)$ using some parametric smoother and then consider sets on the form $\{ \mv{s}; p_i^\tau \geq 1-\rho \}$ where $p_i^\tau$ are the smoothed positive excursion probabilities. 

\begin{defn}[Two-parameter smoothing family]
Let $p_i^\tau$ be the smoothed marginal positive $u$ excursion probabilities, using a circular averaging filter with radius $\tau$. A two-parameter family for the positive and negative $u$ excursion sets is then given by 
  \begin{align*}
D_2^+(\tau,\rho) &= \{ \mv{s}; p_i^\tau \geq 1-\rho \},
\\
D_2^-(\tau,\rho) &= \{ \mv{s}; p_i^\tau \leq \rho \}.
  \end{align*}
\end{defn}
The parameter $\tau$ determines how close $p_i^\tau$ is to the original excursion probabilities. For $\tau=0$, no smoothing is done and for a general $\tau$, $p_i^{\tau}$ is equal to the average of the marginal excursion probabilities in the disk with radius $\tau$ centered at $\mv{s}_i$. As $\tau$ increases $p_i^{\tau}$ becomes smoother and approaches a constant function equal to the average excursion probability. One could also use other types of parametric smoothers instead of the simple averaging filter. 

Using the two-parameter families requires a modification to Algorithm \ref{alg2}, resulting in a slightly more computationally demanding method.

\begin{alg}[Calculating excursion sets using a two-parameter family]\label{alg2}
Assume that the model parameters $\mv{\theta}$ are known and that the posterior distribution $\pi(\mv{x}|\mv{y},\mv{\theta})$ is Gaussian. Further assume that $D(\nu, \rho)$ is a parametric family for the possible excursion sets, such that $D(\nu,\rho_1) \subseteq D(\nu,\rho_2)$ if $\rho_1 < \rho_2$ for a fixed $\nu$. The following strategy is then used to calculate $\exset{u,\alpha}{+}$.
\begin{enumerate}
\item Choose a suitable (sequential) integration method for the problem.
\item Select a suitable one-dimensional optimization strategy.
\item Do optimization of the size of $D(\nu,\bullet)$ over $\nu$:
\begin{itemize}
\item For the current value of $\nu$, reorder the nodes to the order they will be added to the excursion set when the parameter $\rho$ is increased.
\item sequentially add nodes to the set $D$ according to the ordering given above and in each step update the probability $\pP(D\subseteq A_u^+(\mv{x}))$. Stop as soon as this probability falls below $1-\alpha$.
\item return the last set $D$ for which $\pP(D\subseteq A_u^+(\mv{x}))\geq 1-\alpha$.
\end{itemize}
\item $\exset{u,\alpha}{+}$ is given by the largest set $D$ found in the optimization over $\nu$.
\end{enumerate}
\end{alg}

The optimization can in this case be done using a Golden section search or a similar fast optimization procedure for one-dimensional problems. Algorithm~\ref{alg2} can also be used to estimate uncertainty regions for contour curves by using the following two-parameter family for the pair of level avoiding sets.
\begin{defn}[Parametric family for level avoiding sets]\label{def:paravoiding}
Let $D_1^+(\rho_1)$ and $D_1^-(\rho_2)$ be given by Definition \ref{def:parametric0}. A two-parameter family for the pair of level avoiding sets is obtained as $(D_1^+(\rho_1), D_1^-(\rho_2))$. A one-parameter family is obtained by requiring that $\rho_1=\rho_2 = \rho$.
\end{defn}
The one-parameter family in Definition~\ref{def:paravoiding} can be used in Algorithm~\ref{alg1} to estimate level avoiding sets and uncertainty regions for contour curves without having to use the more computationally expensive Algorithm \ref{alg2}.

\subsubsection{Domain bounds and reorderings}
In the case of a GMRF posterior, it is desirable to make the Cholesky factor of the precision matrix as sparse as possible, because it reduces the number of floating point calculations that have to be done and reduces the error of the estimator. Reordering the nodes according to a parametric family does not guarantee good sparsity of the Cholesky factor, but the reordering can be improved by finding upper and lower bounds for the region.

The simplest upper bound for the region is to use 
\begin{align*}
U_1 &= \{\mv{s}: \pP(x(\mv{s})>u)\geq1-\alpha\},
\end{align*}
which is calculated using only the marginal probabilities, and which is the largest region $D$ if $x(\mv{s})$ is a perfectly correlated field. The domain $D$ cannot contain any locations $\mv{s}$ which are not in $U_1$ because all points not in $U_1$ have marginal probabilities lower than $1-\alpha$ of exceeding the level $u$.

A simple lower bound for the region is obtained using Boole's inequality as
\begin{align*}
L_1 &=  \{\mv{\mv{s}}: \pP(x(\mv{s})>u)\geq1-\alpha/n\}
\end{align*}
where $n$ is the number of points in the discretization of the domain. In terms of multiple hypothesis testing, this lower bound is obtained from the classical Bonferroni correction method and an improved lower bound can be obtained using the Holm-Bonferroni method \citep{holm79} as
\begin{align*}
L_2 &=  \{\mv{s}: p_{(k)}>1-\alpha/k\}
\end{align*}
where $p_{(k)}$ is the $k$th largest probability in the set $\{\pP(x(\mv{s}_i)>u), i=1,\ldots,n\}$. If the stochastic variables $x(\mv{s}_i)$ are independent, $L_2$ is the largest domain $D$. If the variables are not independent or perfectly correlated, one has $L_2\subset D \subset U_1$.

The nodes can now be categorized into three classes, the first class contains the nodes included in the lower bound $L_2$, the second class contains the nodes in the set $U_1\setminus L_2$ and the third class contains all other nodes. 
Since one knows that all nodes in $L_2$ will be included in $D$, these can be reordered to maximize the sparsity of the Cholesky factor, for example using an approximate minimum degree permutation. The nodes in the second class are then added in the order determined by the parametric family. Finally, since the nodes in the third class will not be included in the domain, these can be reordered to maximize the sparsity or integrated out of the posterior distribution. Making the bounds more precise will improve the sparsity of the problem and therefore reduce the Monte-Carlo error and the computational complexity.

\subsection{Probability calculations for the latent Gaussian setting}\label{sec:inlaprob}
In practice, we cannot use the computations from the previous sections directly unless we are in a purely Gaussian setting with known parameters. In the latent Gaussian setting with posterior \eqref{eq:posterior}, the method has to be modified. Since this is a latent Gaussian setting, Integrated Nested Laplace Approximations (INLA) \citep{rue09} is a good alternative for estimating the posterior distributions $\pi(\mv{x}|\mv{y},\mv{\theta})$ and $\pi(\mv{\theta}|\mv{y})$; however, the choice of method to use for estimating these distributions does not affect the excursion set estimation so any appropriate method can be used to estimate these distributions. Given estimates of the posterior distributions, we propose three methods for calculating the excursion probabilities, assuming that $\pi(\mv{x}|\mv{y},\mv{\theta})$ is Gaussian:

\begin{description}
\item[EB:] (Empirical Bayes) Ignore the parameter uncertainty and calculate the probability conditionally on a parameter estimate. That is, estimate the excursion sets under the conditional posterior $\pi(\mv{x}|\mv{y},\mv{\theta}_0)$ where $\mv{\theta}_0$ for example is the maximum a posteriori estimate or the maximum likelihood estimate of $\mv{\theta}$. 
\item[QC:] (Quantile Correction) Do a correction to the Gaussian probability calculations based only on the marginal posteriors in the following way. For each $i$ use the marginal posterior to calculate $\pP(x_i>a_i|\mv{y})$ and $\pP(x_i<b_i|\mv{y})$ and calculate $\tilde{a}_i$ and $\tilde{b}_i$ so that $\pP(z_i>\tilde{a}_i|\mv{y},\mv{\theta}_0) = \pP(x_i>a_i|\mv{y})$ and $\pP(z_i<\tilde{b}_i|\mv{y},\mv{\theta}_0) = \pP(x_i<b_i|\mv{y})$. An estimate of the probability is then given by $I(\tilde{\mv{a}},\tilde{\mv{b}},\mv{Q}(\mv{\theta}_0))$, where $\mv{Q}(\mv{\theta}_0)$ is the posterior precision matrix for $\mv{x}$ given the estimated parameters.

\item[NI:] (Numerical Integration) Numerically approximate the excursion probability by approximating the integral in \eqref{eq:posterior} as
\begin{equation*}
\pP(\mv{a}<\mv{x}<\mv{b}|\mv{y}) = \pE(\pP(\mv{a}<\mv{x}<\mv{b}|\mv{y},\mv{\theta})) \approx \sum_{i=1}^k w_i\pP(\mv{a}<\mv{x}<\mv{b}|\mv{y},\mv{\theta}_i)
\end{equation*}
where the configuration of the points $\mv{\theta}_i$ in the hyper parameter space can, for example, be chosen as in the INLA method \citep[see][]{rue09} and the weights $w_i$ are chosen proportional to $\pi(\mv{\theta}_i|\mv{y})$. 
\end{description}

The EB method is the simplest, and may be sufficient in many situations. The QC method is based on correcting the limits of the integral so that the probability would be correct if the $x_i$'s were independent. This method is as easy to implement as the EB method and should perform better in most scenarios. Finally, the NI method is $k$ times more computationally demanding as the probability has to be calculated for each parameter configuration $\mv{\theta}_i$, but should also be the most exact method. If the number of parameters is small one can often obtain accurate results with only a few parameter configurations, but the accuracy of the estimator will depend on how these configurations are chosen.

A second modification is required if the conditional posterior \mbox{$\pi(\mv{x}|\mv{y},\mv{\theta}_0)$} is not Gaussian. The simplest solution to this problem is to do a Gaussian approximation $\tilde{\pi}_G(\mv{x}|\mv{y},\mv{\theta}_0)$, for example using Laplace approximations or simplified Laplace approximations as suggested by \cite{rue09}. If a Gaussian approximation is not sufficient, the sequential integration method has to be modified, and how to do this will depend on the posterior distribution. For example, \cite{Genz09} outline how the quasi Monte Carlo methods can be extended to $t$-distributions, and the GHK-based particle filter method can be extended to other types of distributions as well. However, it is outside the scope of this article to go into details on how to handle the non-Gaussian cases and we therefore leave this for future research.

\section{Tests on simulated data}\label{sec:simulations}
In this section, three examples using simulated data are presented to illustrate the methods and test their accuracy. In the first example, we look at a problem in one dimension with known model parameters, where a latent Gaussian process with an exponential covariance is observed under Gaussian measurement noise. 
In the second example, we compare the different parametric families for contour uncertainty sets for a model in two dimensions with known parameters, where a latent Gaussian Mat\'{e}rn field is observed under Gaussian measurement noise. 
In the third example, the same spatial model setup is used, but this time the model parameters are estimated from data and the three methods for handling the full posterior distribution are compared.

\subsection{Example 1: 1d Gaussian data with known parameters}\label{sec:example1}
\begin{figure}[!t]
\begin{center}
(a)\\
\resizebox{0.8\linewidth}{!}{\input{figs/fig_ex11.tex}
\includegraphics[bb= 30 551 365 690,clip=]{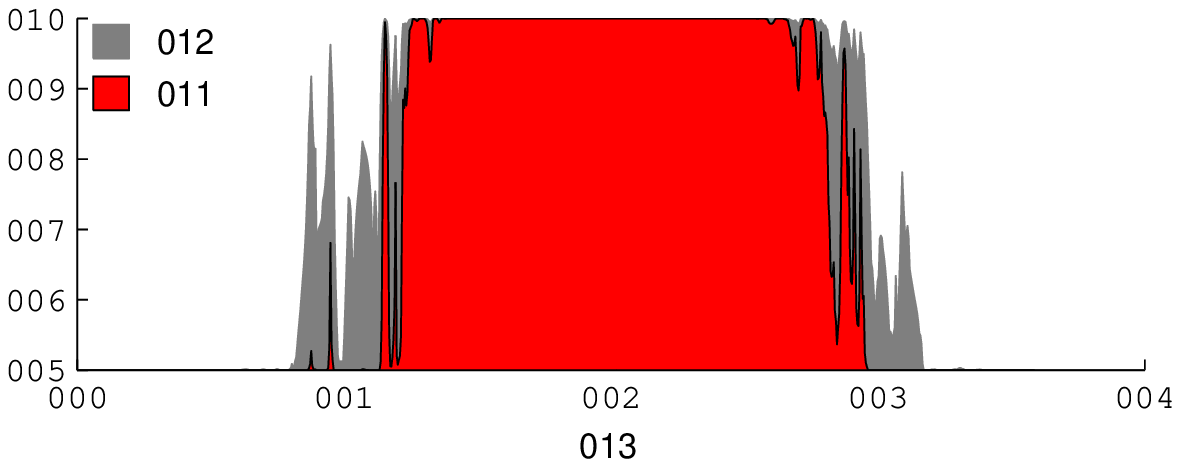}}\\
(b)\\
\resizebox{0.8\linewidth}{!}{\input{figs/fig_ex12.tex}
\includegraphics[bb= 30 551 365 690,clip=]{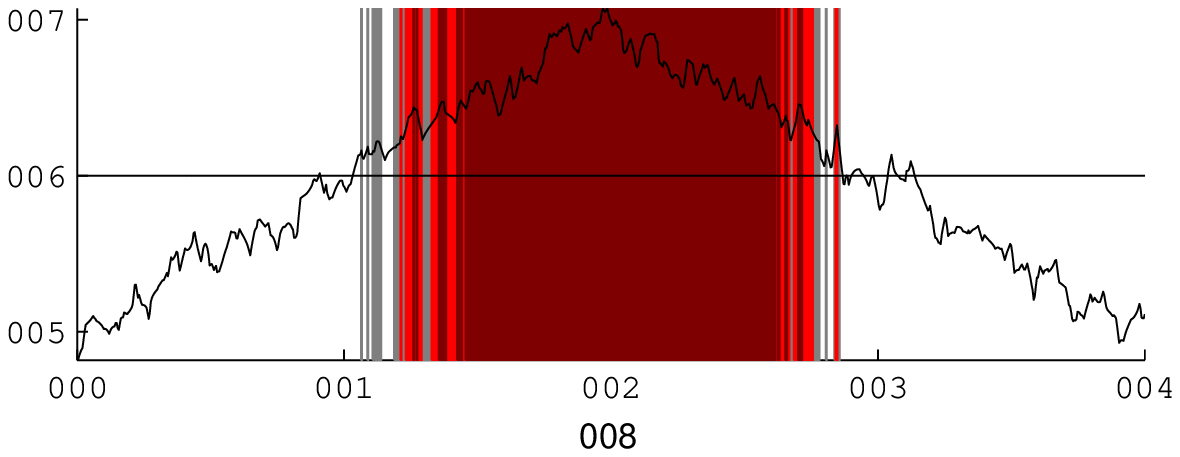}}
\end{center}
\vspace{-0.8cm}
\caption{Results from Example 1. Panel (a) shows the excursion function $F_0^+(s)$ (red) and the marginal excursion probabilities $p(s) = \pP(x(s)>0)$ (grey). Panel~(b) shows $\exset{0,0.05}{+}(x)$ in red, obtained as $A_{0.95}^+(F_0^+)$. The grey area shows $A_{0.95}^+(p)$, which is the upper bound $U_1$, and the dark red set is the lower bound $L_2$. The black curve is the kriging estimate of $x(s)$.}
\label{fig:ex1}
\end{figure}

We begin with a simple one-dimensional example to illustrate the different sets we have previously defined. Let $x(s)$, $s\in [0,2 ]$ be a Gaussian process with an exponential covariance function with scaling parameter $\lambda=1$ and mean 
\begin{equation*}
\mu(s) = 
\begin{cases} 
s-0.5 &\mbox{if } s < 1 \\ 
1.5 -s & \mbox{if } s \geq 1. 
\end{cases} 
\end{equation*}
We generate a trajectory from the model and observe it at $500$ locations $s_1, \ldots, s_n$ drawn at random in the interval under Gaussian measurement noise, giving us observations $y_i = x(s_i) + \vep_i$ where $\vep_i \sim \pN(0,\sigma^2)$. We do spatial prediction (kriging) to $1000$ equally spaced locations in the interval given the parameters $\mv{\theta}$ and the measurements $\mv{y}$, and then estimate the positive $0$-excursion function $F_0^+(s)$ using the parametric family $D_1^+(0,\rho)$. In Figure \ref{fig:ex1}, Panel (a), $F_0^+(s)$ is shown in red together with the marginal excursion probabilities $\pP(x(s)>0)$ in grey.

\begin{figure}[t]
\begin{center}
\begin{minipage}[b]{0.48\linewidth}
\centering
(a)\\
\input{figs/fig_ex13.tex}
\resizebox{\linewidth}{!}{\includegraphics[bb=10 551 198 665,clip=]{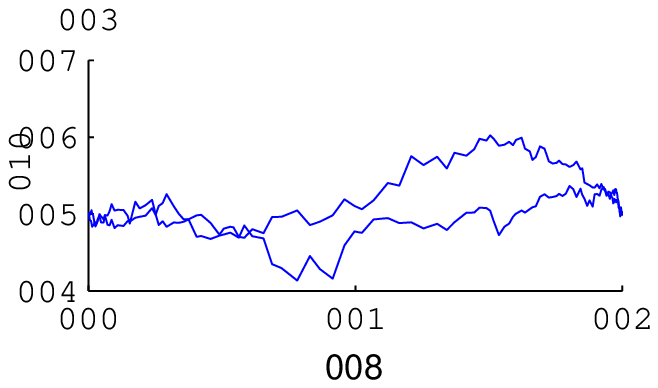}}
\end{minipage}
\begin{minipage}[b]{0.48\linewidth}
\centering
(b)\\
\input{figs/fig_ex14.tex}
\resizebox{\linewidth}{!}{\includegraphics[bb=10 551 198 665,clip=]{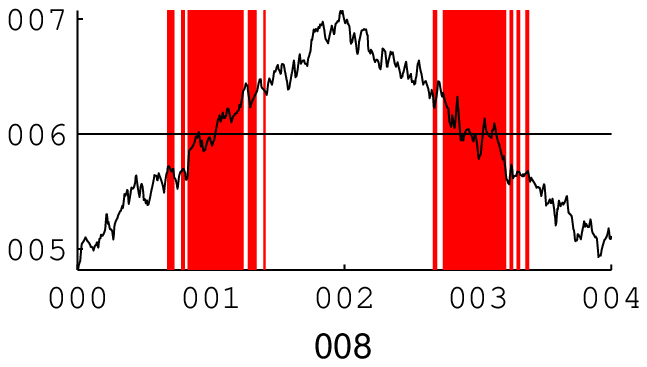}}
\end{minipage}
\end{center}
\vspace{-0.8cm}
\caption{Results from Example 1. Panel (a) shows two estimates of $1-\alpha - \hat{p}(\alpha)$ as a function of $1-\alpha$. $\hat{p}(\alpha)$ is an estimate of $\pP(x(\mv{s})>0, \mv{s}\in \exset{0,\alpha}{+}(x))$ based on Monte-Carlo simulation of $x(\mv{s})$, which should be close to $1-\alpha$ if $\exset{0,\alpha}{+}(x)$ is correctly estimated. The two curves show two results for two different Monte-Carlo simulations. Panel (b) shows the estimated contour uncertainty set $M_{0,0.05}^c(x)$.}
\label{fig:ex12}
\end{figure}

By the definition of $F_0^+(s)$, the positive $0$-excursion set $\exset{0,\alpha}{+}(x)$, is obtained by calculating the $1-\alpha$ excursion set of the function $F_0^+(s)$, and this set is shown for $\alpha=0.05$ in red in Figure \ref{fig:ex1}, Panel (b). The grey set shows the upper bound $U_1$, which is the set where $\pP(x(s)>0)\geq1-\alpha$, and the dark red set shows the Holm-Bonferroni lower bound $L_2$. The black curve shows the kriging estimate of the process given the data. Note that the grey and red sets are obtained as excursion sets of the grey and red functions in Panel (a), and also note that $L_2\subset \exset{0,\alpha}{+}(x) \subset U_1$. 

We now want to verify that the estimated sets $\exset{0,\alpha}{+}(x)$ have the correct excursion probability, that is, that $\pP(x(s)>0,s\in \exset{0,\alpha}{+}(x))=1-\alpha$. To that end, draw $N$ samples, $x_1(s), \ldots, x_N(s)$ from $\pi(x|\mv{y},\mv{\theta})$, count the number of samples for which $\inf\{x(s),s\in \exset{0,\alpha}{+}(x)\} \geq 0$, and denote this number by $N_s$. Further let $\hat{p}(\alpha)$ denote the proportion of samples, $N_s/N$, that satisfies the requirement. If $\exset{0,\alpha}{+}(x)$ is correctly estimated, $\hat{p}(\alpha)$ should be close to $1-\alpha$. In Figure \ref{fig:ex12}, Panel~(a), the difference $1-\alpha-\hat{p}(\alpha)$ is shown as a function of $1-\alpha$. The difference is calculated twice, using two different estimates $\hat{p}(\alpha)$, each based on $N=50000$ samples. As can be seen in the figure, the difference is very small for all values of $\alpha$, and the difference that can be seen is mostly due to the Monte-Carlo error in the estimation of $\hat{p}(\alpha)$, which has nothing to do with the accuracy of the method. Thus the sets $\exset{0,\alpha}{+}$ indeed have the correct excursion probabilities. 

Finally in Figure \ref{fig:ex12}, Panel (b), the $0$-contour uncertainty region $M_{0,0.05}^c(x)$ is shown in red and the kriging estimate of $x(s)$ is again shown in black. The set was estimated using the two-parameter family for level avoidance sets from Definition \ref{def:paravoiding} and Algorithm \ref{alg2}. The complement of this set is the union of the level avoiding sets $(M_{0,0.05}^-(x),M_{0,0.05}^+(x))$, which is the largest pair of sets $(D^+,D^-)$ satisfying $\pP(D^- \subseteq A_u^-(x),\, D^+ \subseteq A_u^+(x)) \geq 0.95$.

\subsection{Example 2: 2d Gaussian data with known parameters}
In this example, we change to a spatial model to test the parametric families for contour sets. Let $x(\mv{s})$, $\mv{s}\in [0,10] \times [0,10]$, be a Gaussian field with a constant mean $\mu=0$ and a Mat\'{e}rn covariance function 
\begin{equation}\label{paperE:eq:matern}
C(\|\mv{h}\|) = \frac{2^{1-\nu}\phi^2}{(4\pi)^{\frac{d}{2}}\Gamma(\nu + \frac{d}{2})\kappa^{2\nu}}(\kappa\|\mv{h}\|)^{\nu}K_{\nu}(\kappa\|\mv{h}\|),
\end{equation}
where $\nu$ is a shape parameter, $\kappa^2$ a scale parameter, $\phi^2$ a variance parameter, $K_{\nu}$ is a modified Bessel function of the second kind of order $\nu>0$, and $\|\cdot\|$ denotes the Euclidean spatial distance. We use the SPDE representation by \cite{lindgren10} of the field using a triangulation based on an $80 \times 80$ regular lattice in the region. The representation is a piecewise linear approximation $x(\mv{s}) \approx \sum_i x_i \varphi_i(\mv{s})$ of the field using $6400$ piecewise linear functions $\varphi_i(\mv{s})$, each centered at one of the nodes in the lattice. The advantage with this representation is that it allows us to do all calculations using the weights $\mv{x}$ of the basis expansion, which form a Gaussian Markov random field. 

\begin{figure}[t]
\begin{center}
\begin{minipage}[b]{0.27\linewidth}
\centering
(a)\\
\includegraphics[width=\linewidth,bb=140 227 496 583,clip=]{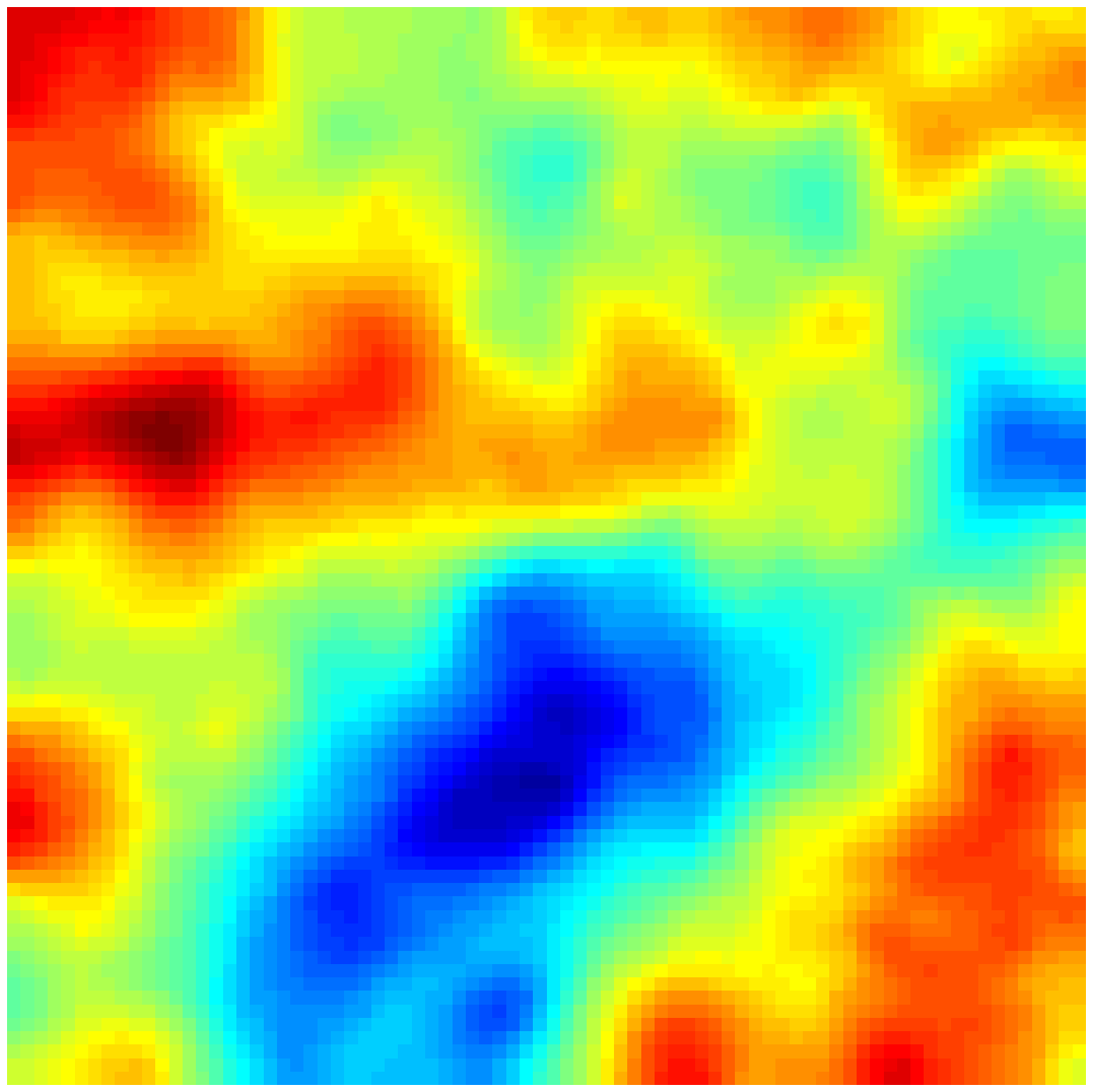}
\end{minipage}
\begin{minipage}[b]{0.27\linewidth}
\centering
(b)\\
\includegraphics[width=\linewidth,bb=140 227 496 583,clip=]{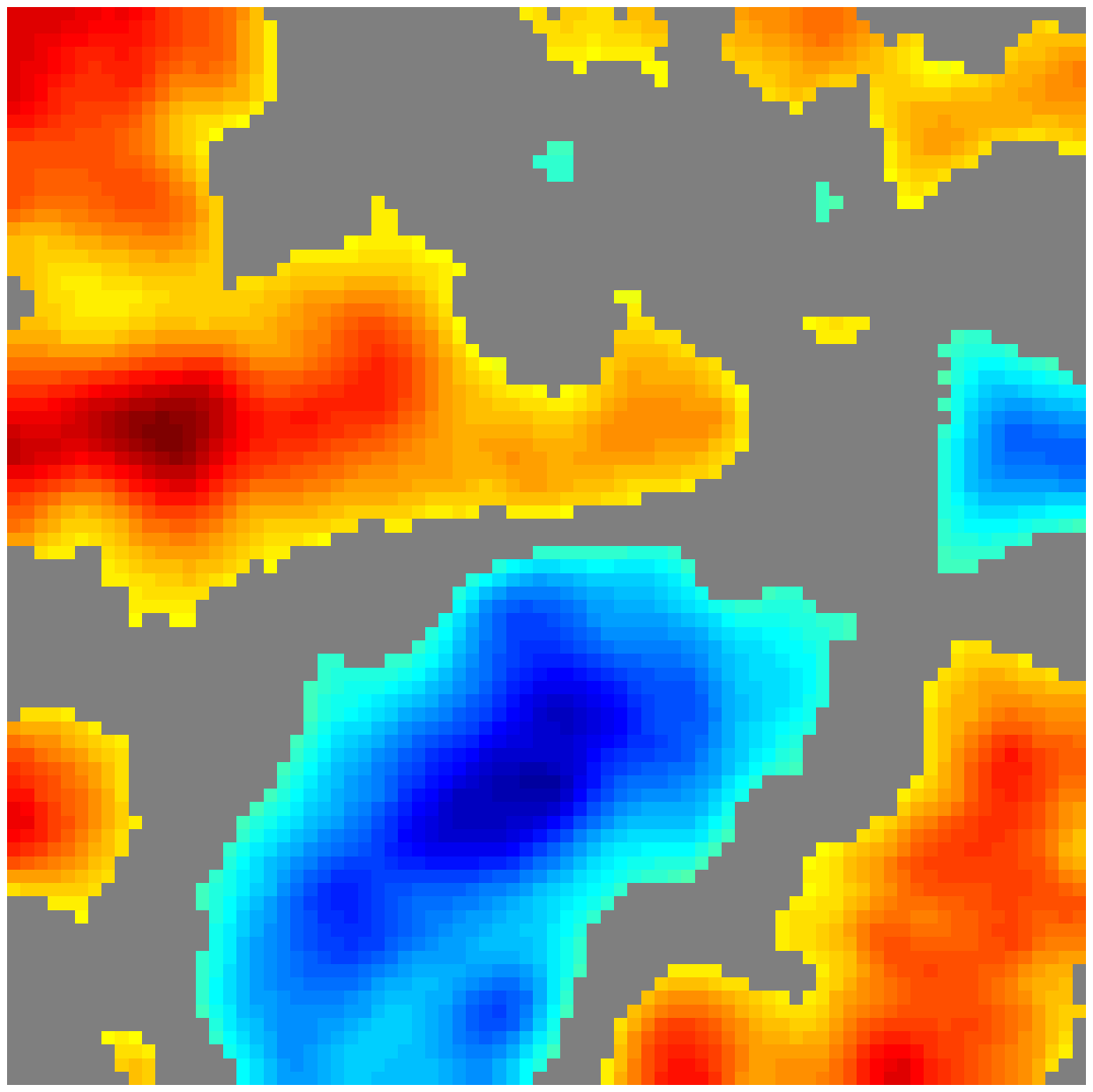}
\end{minipage}
\begin{minipage}[b]{0.05\linewidth}
\centering
\quad\\
\input{figs/fig_ex33.tex}
\includegraphics[height=49mm,width=11mm,bb= 65 2860 102 3063,clip=]{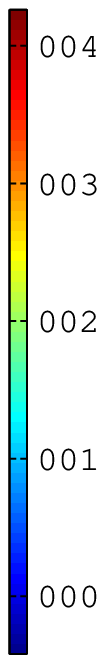}%
\end{minipage}
\end{center}
\vspace{-0.8cm}
\caption{Results from Example 2. Panel (a) shows a kriging estimate and Panel (b) shows the same estimate where the corresponding estimated contour uncertanty set $M_{0,0.05}^c(x)$ is superimposed in grey.}
\label{fig:ex31}
\end{figure}

We set $\nu = \phi = 1$, and $\kappa^2=0.5$, and generate a sample of the field and measure it at $1000$ locations in the square, chosen at random, under Gaussian measurement noise, giving us observations $y_i = x(\mv{s}_i) + \vep_i$ where $\vep_i \sim \pN(0,\sigma^2)$ and $\sigma=0.1$. The posterior estimate (kriging) of $\mv{x}|\mv{y}$ can be seen in Figure \ref{fig:ex31}, Panel (a), and the uncertainty region $M_{0,0.05}^c(x)$ for the $0$-contour can be seen in Panel (b). In this case the $M_{0,0.05}^c(x)$ was estimated using the one-parameter family in Definition \ref{def:paravoiding}, and it is now of interest to test how much is gained by using the two-parameter family from the same definition instead. 

To that end, we generate $50$ data sets using the same setup, and for each data set estimate $M_{0,0.05}^c(x)$, first using the one-parameter family \mbox{$(D_1^+(\rho),D_1^-(\rho))$}, and then using the more general two-parameter family \mbox{$(D_1^+(\rho_1),D_1^-(\rho_2))$}. Since the one-parameter family is a special case of the two-parameter family where ${\rho_1=\rho_2=\rho}$, the contour sets estimated with the two-parameter family should always be smaller than the one-parameter sets. However, using the two-parameter family, the estimated sets are on average only $0.2\%$ smaller than if the one-parameter family is used, so in this case it is arguably not worth the extra computational effort to use the two-parameter family, although for other levels $u$, or other latent models, the difference might be larger.

\subsection{Example 3: 2d Gaussian data with unknown parameters}
In this example we compare the three methods,described in Section \ref{sec:inlaprob}, for handling the full posterior distribution \eqref{eq:posterior} in the calculations. The same Gaussian Mat\'{e}rn model is used as in Example 2, with the difference that we now also estimate the parameters from the data.  

\begin{figure}[!t]
\vspace{-0.5cm}
\begin{center}
\hspace{-0.2cm}\begin{minipage}[b]{0.27\linewidth}
\centering
\phantom{Q}EB\phantom{Q}\\
\includegraphics[width=\linewidth,bb=140 227 496 583,clip=]{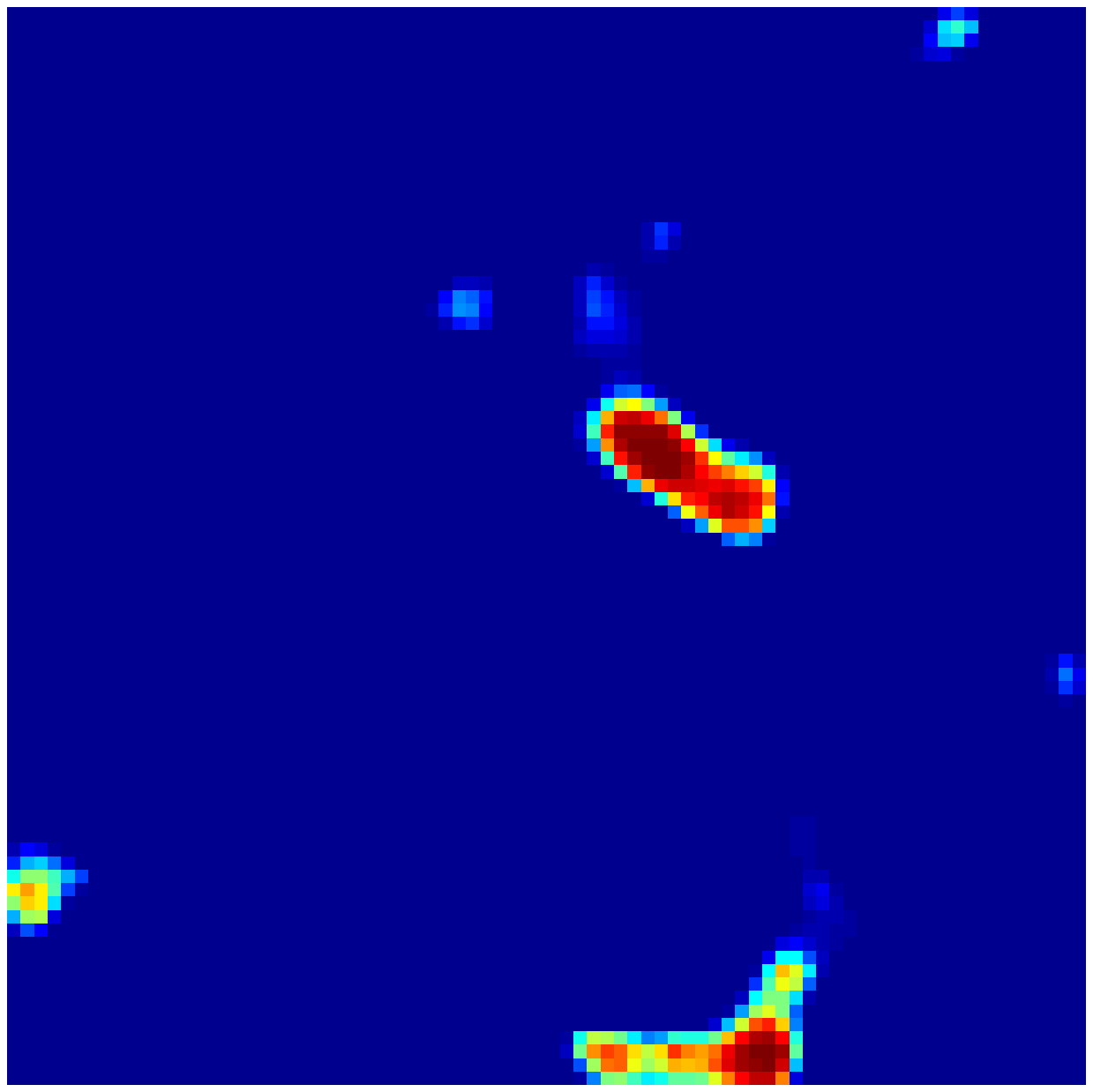}%
\end{minipage}
\begin{minipage}[b]{0.27\linewidth}
\centering
QC\\
\includegraphics[width=\linewidth,bb=140 227 496 583,clip=]{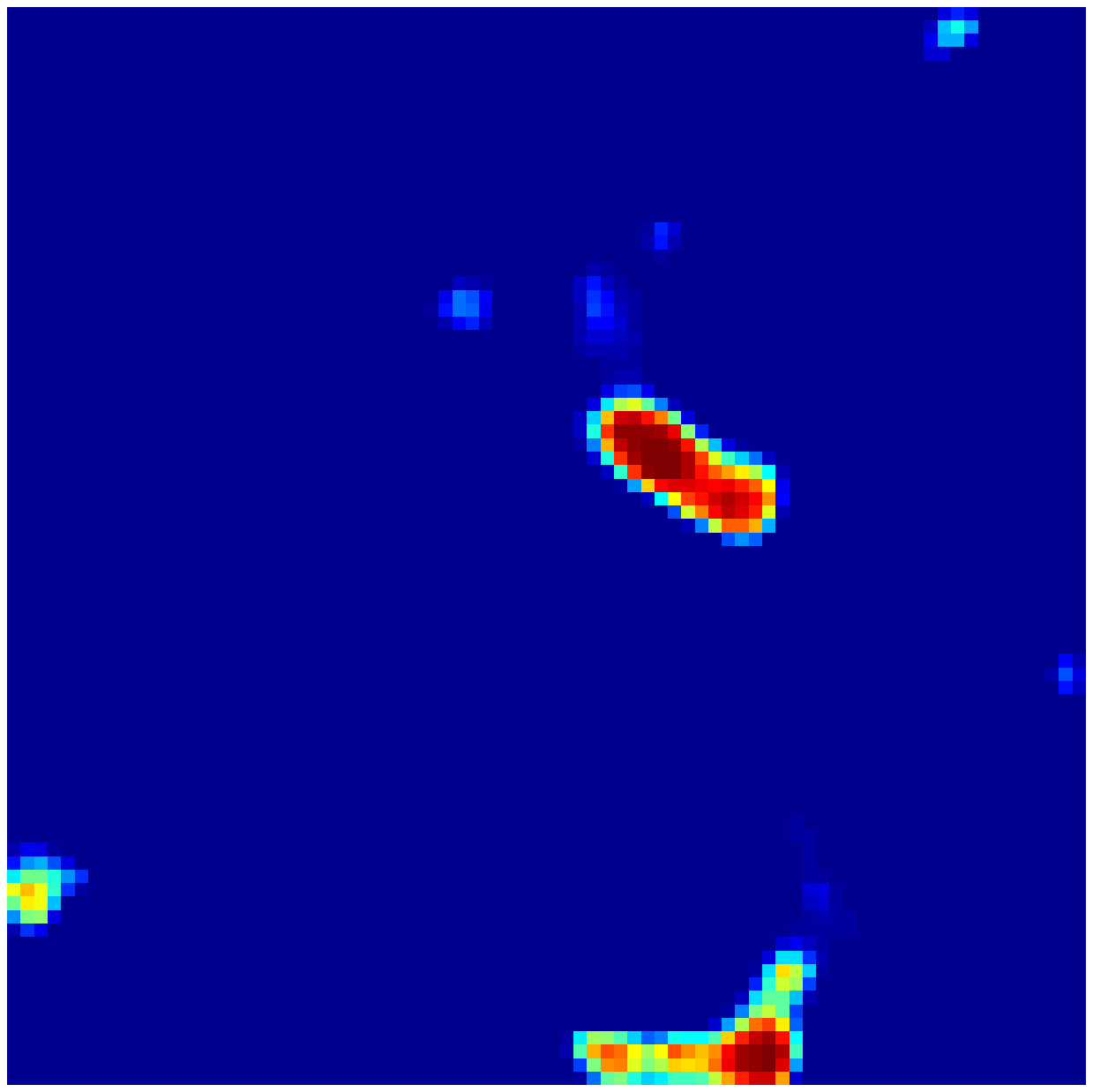}%
\end{minipage}
\begin{minipage}[b]{0.27\linewidth}
\centering
\phantom{Q}NI\phantom{Q}\\
\includegraphics[width=\linewidth,bb=140 227 496 583,clip=]{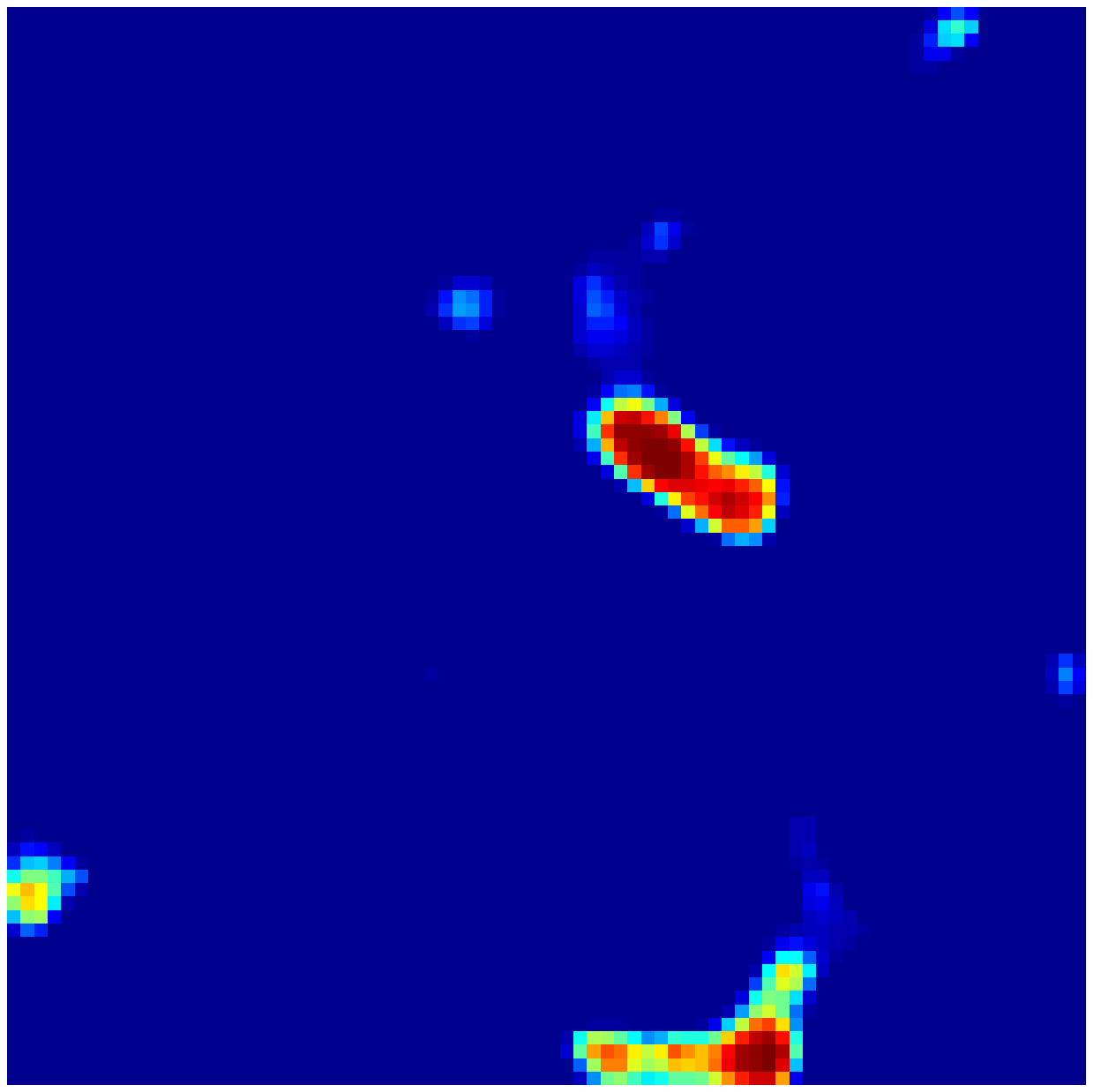}%
\end{minipage}
\begin{minipage}[b]{0.05\linewidth}
\centering
\quad\\
\input{figs/fig_ex24.tex}
\includegraphics[height=49mm,bb=65 2853 102 3006,clip=]{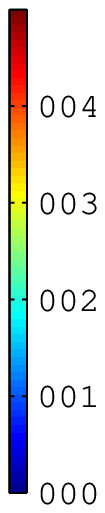}%
\end{minipage}

\hspace{-0.2cm}\begin{minipage}[b]{0.27\linewidth}
\vspace{0.2cm}
\centering
Marginal p-values\\
\includegraphics[width=\linewidth,bb=140 227 496 583,clip=]{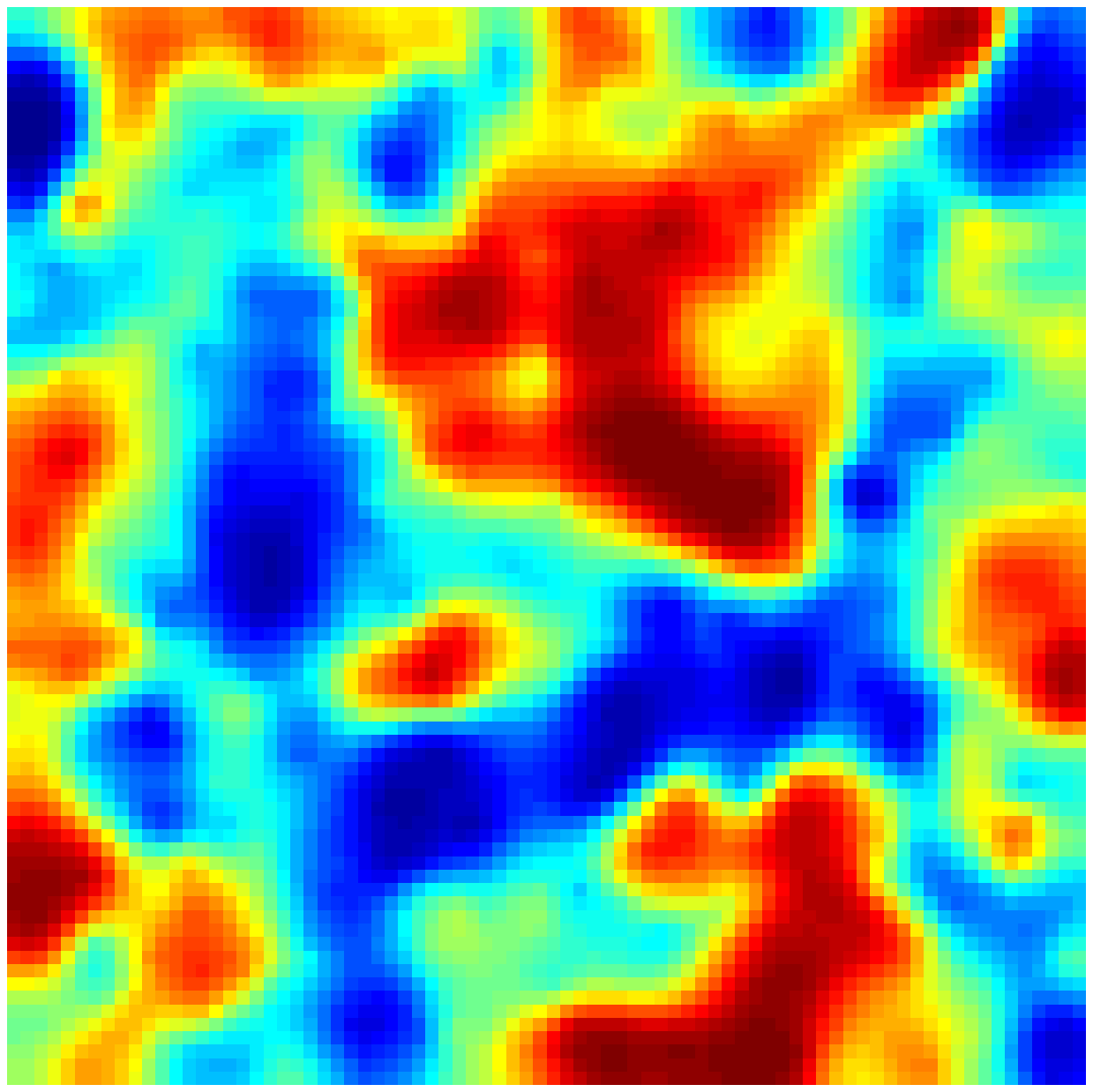}%
\end{minipage}
\begin{minipage}[b]{0.27\linewidth}
\centering
\phantom{p}Excursion set\phantom{p}\\
\includegraphics[width=\linewidth,bb=140 227 496 583,clip=]{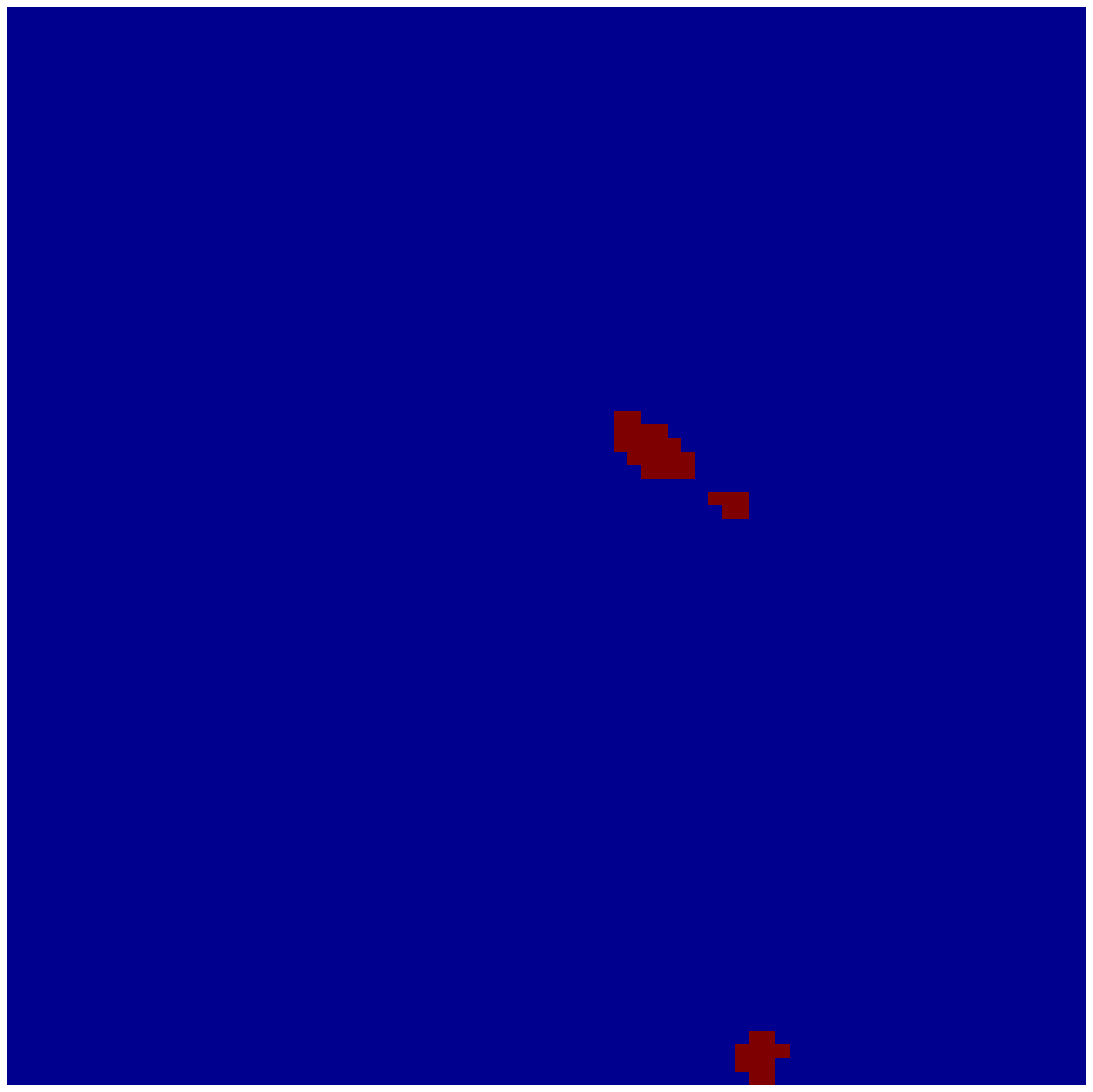}%
\end{minipage}
\begin{minipage}[b]{0.27\linewidth}
\centering
Kriging estimate\\
\includegraphics[width=\linewidth,bb=140 227 496 583,clip=]{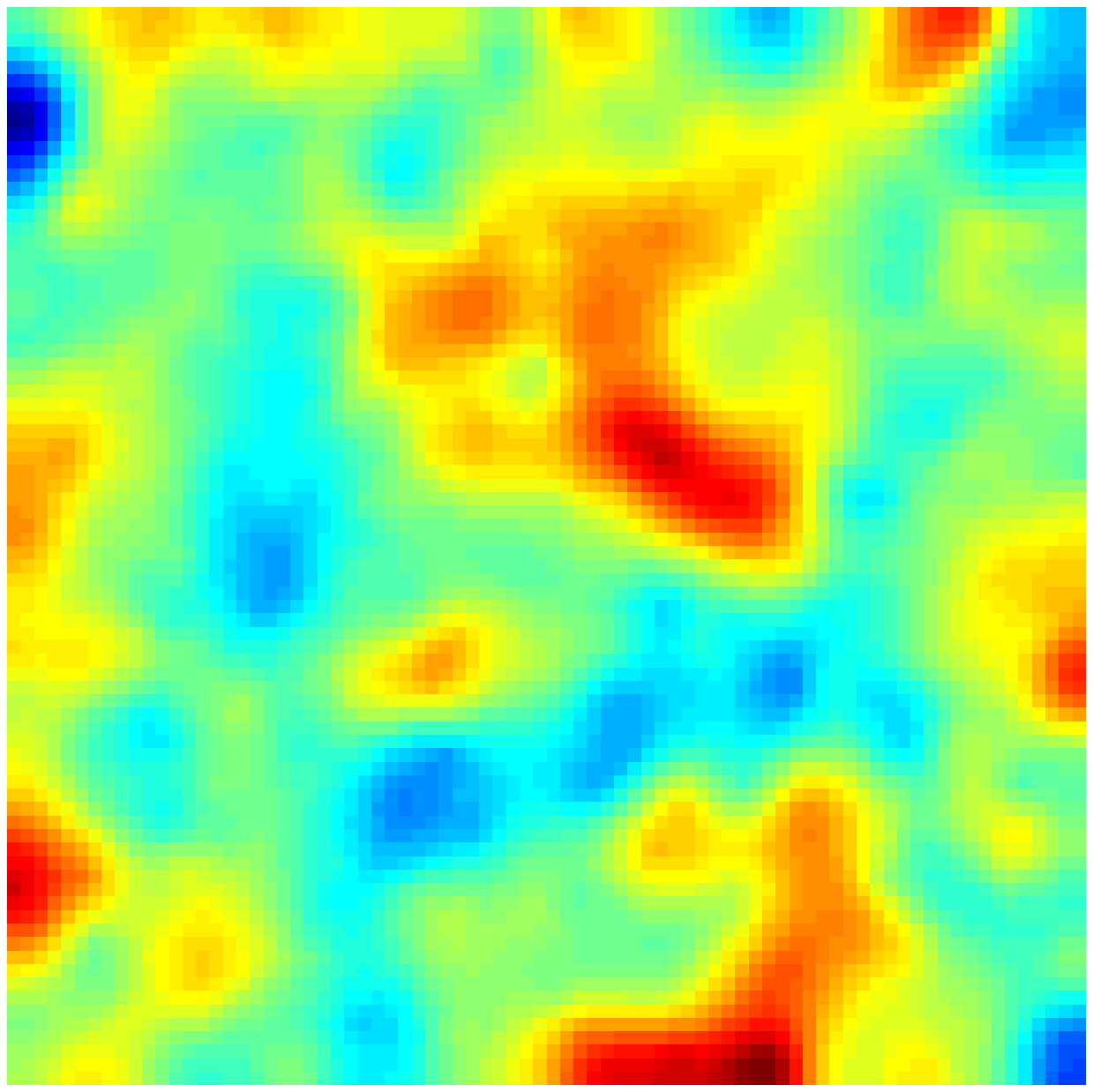}%
\end{minipage}
\begin{minipage}[b]{0.05\linewidth}
\centering
\quad\\
\input{figs/fig_ex28.tex}
\includegraphics[height=49mm,bb=65 2853 102 3006,clip=]{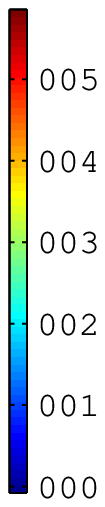}%
\end{minipage}
\end{center}
\vspace{-0.4cm}
\caption{Results from Example 3. In the top row, three estimates of the excursion function can be seen using the EB method (left), the QC method (middle), and the NI method with $15$ parameter configurations (right). In the bottom row, the marginal $p$-values for exceeding the limit can be seen in the left panel, using the same color scale as for the top row. The middle panel shows the set $\exset{0,0.05}{+}(x)$ given by excursion function estimated by the NI method. Finally the right panel shows the kriging estimate of the latent field.}
\label{fig:ex2}
\end{figure}

We set $\nu = \phi = 1$, and $\kappa^2=2$ in the covariance function \eqref{paperE:eq:matern}, and generate a sample of the field and measure it at $1000$ locations in the square, chosen at random, under Gaussian measurement noise, giving us observations $y_i = x(\mv{s}_i) + \vep_i$ where $\vep_i \sim \pN(0,\sigma^2)$ and $\sigma=0.5$. Given the measurements, we estimate the parameters and the marginal posterior distributions using INLA. The posterior estimate (kriging) of $\mv{x}|\mv{y}$ can be seen in the lower right panel of Figure \ref{fig:ex2}, and in the lower left panel, the marginal probabilities $\pP(x(\mv{s})>0|\mv{y})$ are shown.

We now estimate the excursion function $F_0^+(\mv{s})$ using the three different methods described in Section~\ref{sec:inlaprob} and the one-parameter family from Definition~\ref{def:parametric0} for the excursion sets. These can be seen in the upper panels of Figure~\ref{fig:ex2}. Visually it is in this case difficult to see any differences between the three estimates of the excursion function. To compare the accuracy of the estimates we will do a simular comparison to the one performed in Example 1, where Monte-Carlo simulation was used to estimate $\hat{p}(\alpha)$, the proportion of samples satisfying $\inf\{x(\mv{s}),\mv{s}\in \exset{0,\alpha}{+}(x)\} \geq 0$, which should be close to $1-\alpha$ if $\exset{0,\alpha}{+}(x)$ is correct. 

There are three possible sources of errors in this comparison. The first one is the Monte-Carlo error from the estimation of $\hat{p}(\alpha)$, which has nothing to do with the accuracy of the method. The second error is the Monte-Carlo error in the probability estimation when estimating the excursion distribution functions. This error is, however many orders of magnitude smaller in this case. The final error is the approximation error induced by using any of the three methods EB, QC, or NI for handling the full posterior distribution. 

\begin{figure}[t]
\begin{center}
\begin{minipage}[b]{0.49\linewidth}
\centering
(a)\\
\resizebox{\linewidth}{!}{\input{figs/fig_ex2mcmc3.tex}
\includegraphics[bb=5 551 190 665,clip=]{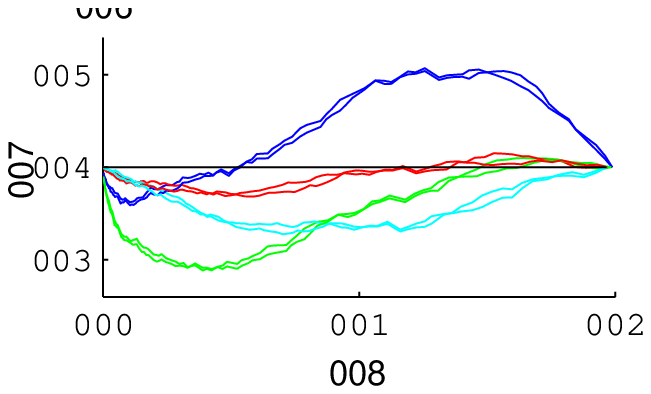}}%
\end{minipage}
\begin{minipage}[b]{0.49\linewidth}
\centering
(b)\\
\resizebox{\linewidth}{!}{\input{figs/fig_ex2mcmc4.tex}
\includegraphics[bb=5 551 190 665,clip=]{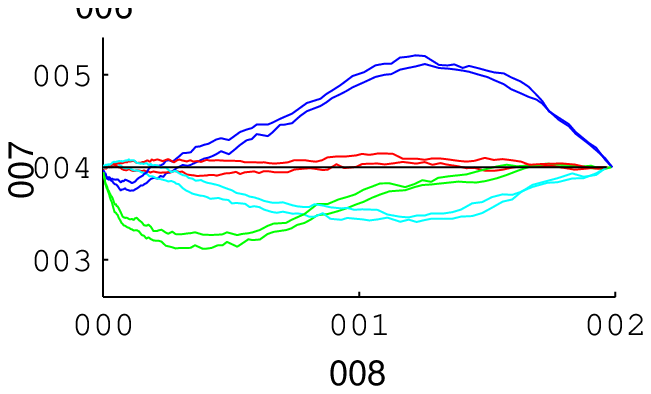}}%
\end{minipage}
\end{center}
\vspace{-0.6cm}
\caption{Results from Example 3 showing the difference \mbox{$1-\alpha - \hat{p}(\alpha)$} as a function of $1-\alpha$ for the different approximation methods, EB (blue), QC (green), NI using $45$ parameter configurations (red), and NI using $15$ parameter configurations (cyan). $\hat{p}(\alpha)$ is an estimate of $\pP(x(\mv{s})>0, \mv{s}\in \exset{0,\alpha}{+}(x))$ based on MCMC simulation of $x(\mv{s})$, which should be close to $1-\alpha$ if $\exset{0,\alpha}{+}(x)$ is correctly estimated. In Panel (a), $\hat{p}(\alpha)$ is estimated using the full posterior distribution, and in Panel~(b), $\hat{p}(\alpha)$ is estimated using the discrete posterior distribution defined by the $45$ parameter configurations from the NI method. The comparison was done twice, with the two different estimates of $\hat{p}(\alpha)$, each based on $20 000$ samples of $x(\mv{s})$, and the curves of the same color shows these two. }
\label{fig:ex2mcmc1}
\end{figure}

To investigate this approximation error, the difference \mbox{$1-\alpha-\hat{p}(\alpha)$} is estimated for the three estimates of $F_0^+(\mv{s})$. First we base the estimate on $20 000$ samples from the posterior $\pi(\mv{x}|\mv{y})$, obtained using the MCMC sampler described in Appendix~\ref{sec:appendixa}. In Figure \ref{fig:ex2mcmc1}, Panel (a), the results can be seen for the EB method (blue), the QC~method (green), the NI method with $k=45$ parameter settings (red), and the NI method with $k=15$ parameter settings (cyan). The comparison was done twice, with two different samples of size $20 000$ when calculating $\hat{p}(\alpha)$, and the curves of the same color show these two and give an indication of the size of the Monte-Carlo error in the comparison. As seen in the figure, the NI method performs best, as expected. 

The error using the NI method comes from the fact that only finitely many points are used in the integration when approximating the posterior distribution for the parameters. That is, the full posterior $\pi(\mv{\theta}|\mv{y})$ is approximated by a discrete distribution with point masses at the parameter configurations $\mv{\theta}_i$ used in the integration, $\pi(\mv{\theta}_i) = w_i$. To verify that this indeed is the source of the error in, we construct a second Monte-Carlo sampler where we instead of sampling $\mv{\theta}$ from the full posterior $\pi(\mv{\theta}|\mv{y})$ sample it from the discrete distribution defined by the $45$ parameter configurations used in the first NI method. Panel (b) in Figure \ref{fig:ex2mcmc1} shows the same comparison as Panel (a) but where $\mv{\theta}$ is sampled from the discrete distribution. As expected, the error for the NI method with $45$ parameter configurations is now smaller. 

\begin{figure}[t]
\begin{center}
\begin{minipage}[b]{0.49\linewidth}
\centering
(a)\\
\resizebox{\linewidth}{!}{\input{figs/fig_ex2mcmc1.tex}
\includegraphics[bb=5 551 190 665,clip=]{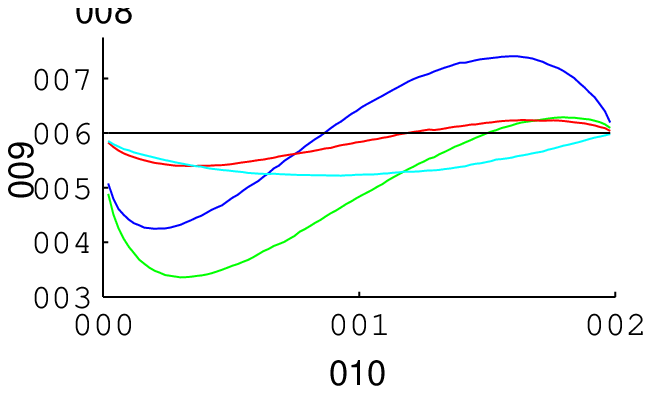}}%
\end{minipage}
\begin{minipage}[b]{0.49\linewidth}
\centering
(b)\\
\resizebox{\linewidth}{!}{\input{figs/fig_ex2mcmc2.tex}
\includegraphics[bb=5 551 190 665,clip=]{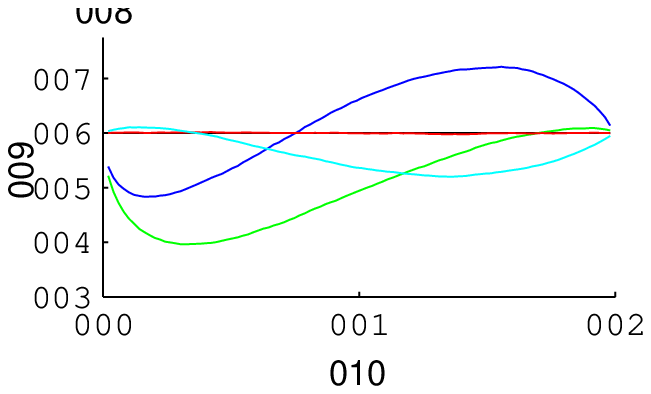}}%
\end{minipage}
\end{center}
\vspace{-0.6cm}
\caption{Results from Example 3 showing the difference $1-\alpha - \hat{p}(\alpha)$ as a function of $1-\alpha$ for the different approximation methods, EB (blue), QC (green), NI using $45$ parameter configurations (red), and NI using $15$ parameter configurations (cyan). The same setup as in Figure \ref{fig:ex2mcmc1} was repeated $50$ times for $50$ different datasets and $60 000$ samples were used when estimating $\hat{p}(\alpha)$. This figure shows the average error of these $50$ runs.}
\label{fig:ex2mcmc2}
\end{figure}

The Monte-Carlo error from estimating $\hat{p}(\alpha)$ is quite large in Figure \ref{fig:ex2mcmc1}, so to get a better understanding of the other errors, a larger study was also performed where the procedure in Figure \ref{fig:ex2mcmc1} was repeated $50$ times for $50$ different simulated data sets, and for each data set $N=60 000$ draws from the posterior was used when estimating $\hat{p}(\alpha)$. The average errors of these $50$ runs can be seen in Figure~\ref{fig:ex2mcmc2}. In Panel (a) the results using samples from the full posterior is shown, and in Panel~(b) the results using the discrete distribution for $\mv{\theta}$ is shown. Note that the red curve is very close to zero in Panel (b), indicating that the error in the NI method mostly depends on choosing the integration points for $\mv{\theta}$ so that they capture the true posterior distribution well. Also note that the QC method performs well for large values of $1-\alpha$, and since one most often is interested in finding the excursion sets for small values of $\alpha$, this method is then a good way of finding such sets with less computational effort than using the NI~method.

\section{Applications}\label{sec:applications}
In this section, we will use the techniques described above in two different applications. In the first, we study air pollution data from Piemonte region in northern Italy and estimate regions where the daily limit for PM$_{10}$ (particulate matter with an aerodynamic diameter of less than 10 $\mu$m) is exceeded. In the second application, we study vegetation index data from the African Sahel and estimate areas that experienced a significant increase in vegetation after the drought period in the early 1980's. In both of these applications the data sets are large, and the Markov structure of the latent Gaussian models has to be used in the calculations.

\subsection{Air pollution data}
High levels of air pollution can be harmful for the ecosystems and the human health. The effects on human health ranges from minor effects to the cardio-respiratory system to premature mortality \citep{Cohen09, cameletti12}. Because of this, environmental agencies have to assess the air quality in order to take proper actions for improving the situation in polluted areas, and an important tool in this process is the ability to produce continuous maps of air pollution. 

A region where the daily limit values fixed by the European Union for human health protection (see EU Council Directive 1999/30/EC) are periodically exceeded is the Piemonte region in northern Italy. Recently, \cite{cameletti12} proposed a statistical model to capture the complex spatio-temporal dynamics of PM$_{10}$ concentration in the region and used it to produce daily maps of PM$_{10}$. They also produced daily maps of exceedance probabilities of the value $50 \mu g/m^3$, which is the value fixed by the European directive 2008/50/EC for the daily mean concentration that cannot be exceeded more than 35 days in a year. These probability maps only considered the marginal excursion probabilities, and no attempts of producing maps of simultaneous exceedance probabilities were made. In the following, we will therefore consider the same model and data but also estimate the excursion functions for the $50 \mu g/m^3$ limit value.

\cite{cameletti12} considers daily PM$_{10}$ data measured at 24 monitoring stations by the Piemonte monitoring network during $182$ days in the period October 2005 - March 2006, the data is provided by Aria Web Regione Piemonte. Denoting the measurements made at location $\mv{s}_i$ at time $t$ by $y(\mv{s}_i,t)$, the following measurement equation is assumed,
\begin{equation}
y(\mv{s}_i,t) = x(\mv{s}_i,t) + \vep(\mv{s}_i,t),
\end{equation}
where $\vep(\mv{s}_i,t)\sim\pN(0,\sigma_{\vep}^2)$ is Gaussian measurement noise, both spatially and temporally uncorrelated, and $x(\mv{s}_i,t)$ is the latent field of true unobserved air pollutions. The latent field is assumed to be on the form
\begin{equation}
x(\mv{s}_i,t) = \sum_{k=1}^p z_k(\mv{s}_i,t)\beta_k + \xi(\mv{s}_i,t),
\end{equation}
where the $p=9$ covariates $z_k$ are used and $\xi$ is a spatio-temporal Gaussian random field. Based on the work of \cite{cameletti11} the following covariates were used:
\begin{inparaenum}[\itshape 1\upshape)]
\item Daily mean wind speed;
\item daily maximum mixing height;
\item daily precipitation;
\item daily mean temperature;
\item daily emissions;
\item altitude;
\item longitude;
\item latitude; and
\item intercept.
\end{inparaenum}
These covariates are provided with hourly temporal resolution on a $4$ km $\times$ $4$ km regular grid by the environmental agency of Piemonte region (Arpa Piemonte), see \cite{Finardi08}. The spatio-temporal process $\xi$ is assumed to follow first order autoregressive dynamics in time with spatially dependent innovations:
\begin{equation}
\xi(\mv{s}_i,t) = a\xi(\mv{s}_i,t-1) + \om(\mv{s}_i,t),
\end{equation}
where $|a|<1$ and $\om(\mv{s}_i,t)$ is a zero-mean temporally independent Gaussian process characterized by the spatio-temporal covariance function
\begin{equation}
\Cov(\om(\mv{s}_i,t_1),\om(\mv{s}_j,t_2)) = \begin{cases}
0 &\mbox{if } t_1 \neq t_2 \\
C(\|\mv{s}_i-\mv{s}_j\|) &\mbox{otherwise,}
\end{cases}
\end{equation}
where $C(\cdot)$ is a Mat\'{e}rn covariance function given by \eqref{paperE:eq:matern}. 
The model parameters and the posterior distribution for the latent field are then estimated using INLA in combination with the SPDE representation of \cite{lindgren10}, see \cite{cameletti12} for details.

\begin{figure}
\begin{center}
\begin{minipage}[b]{0.35\linewidth}
\centering
Marginal probabilities\\
\includegraphics[width=\linewidth]{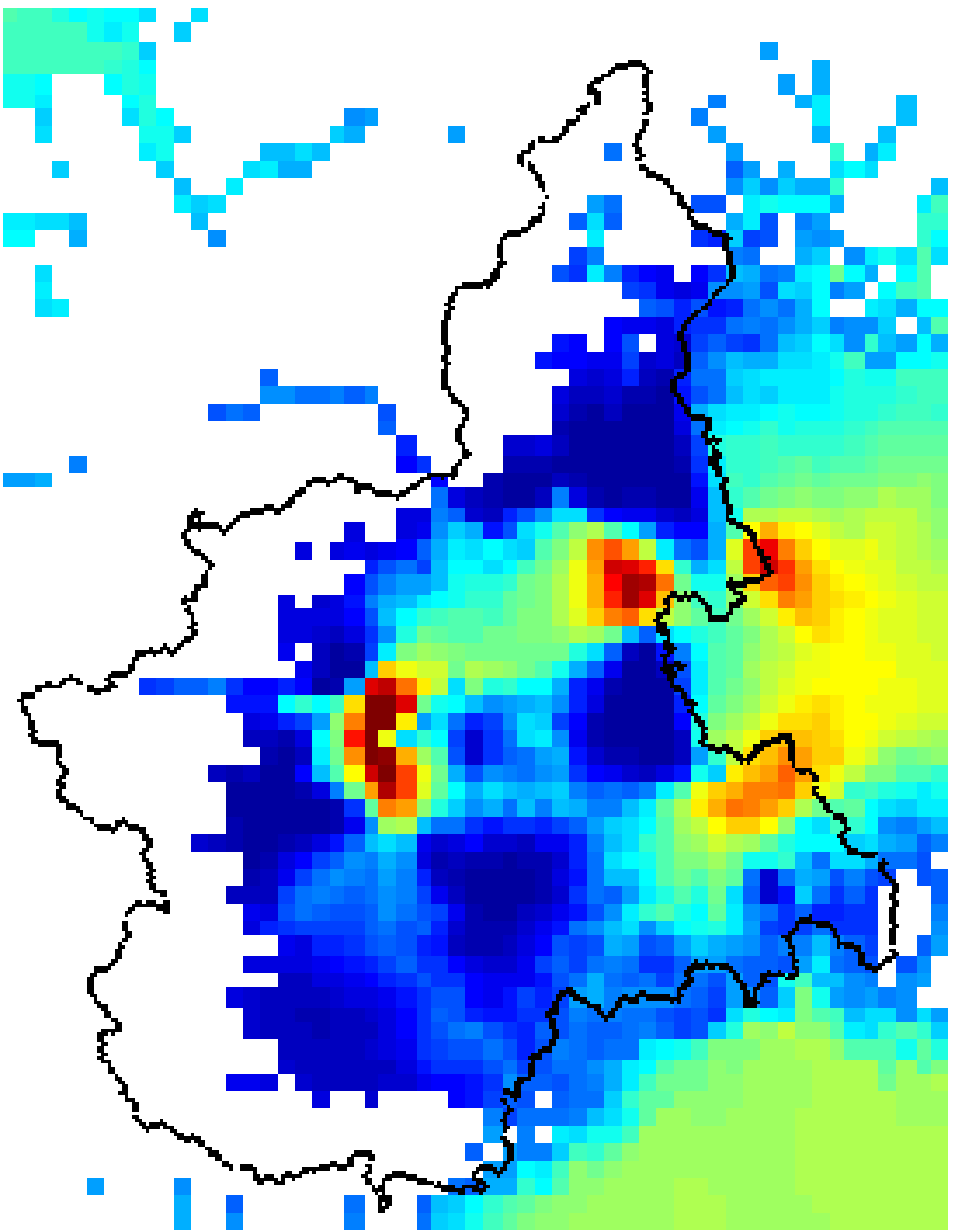}%
\end{minipage}
\begin{minipage}[b]{0.35\linewidth}
\centering
\phantom{p}Excursion function $F_{50}^+(\mv{s})$\phantom{p}\\
\includegraphics[width=\linewidth]{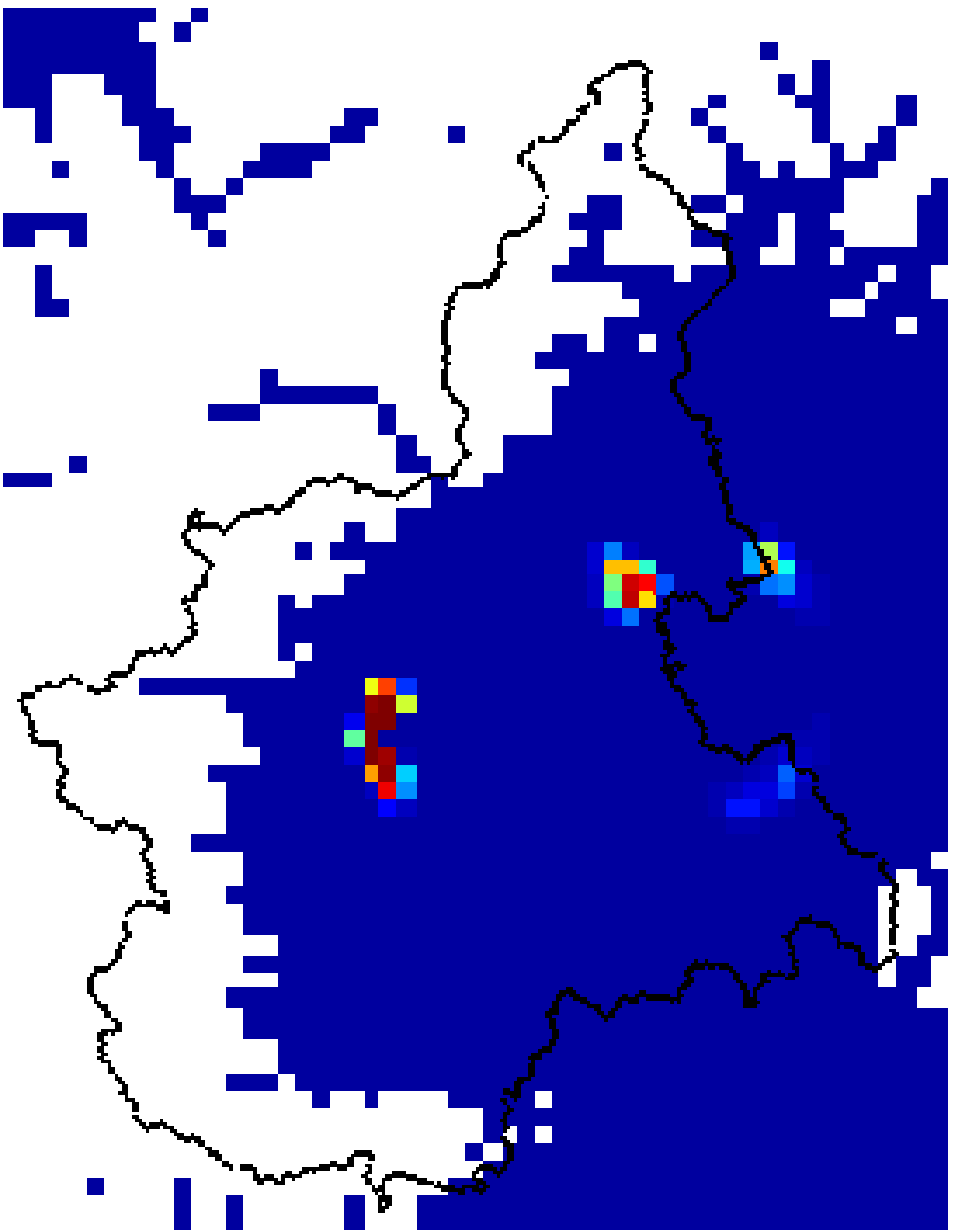}%
\end{minipage}
\begin{minipage}[b]{0.05\linewidth}
\centering
\quad\\
\input{figs/pm10_cbar.tex}
\includegraphics[height=81mm,bb= 65 2860 90 3063,clip=]{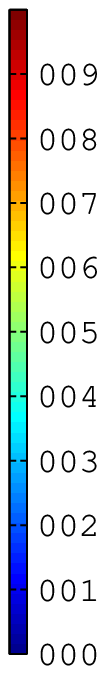}%
\end{minipage}
\end{center}
\vspace{-0.4cm}
\caption{Results from the PM$_{10}$ application for January 30, 2006. A map of the marginal exceedance probabilities for $50 \mu g/m^3$ (left), and the joint excursion distribution function for the level (right).}
\label{fig:pm10}
\end{figure}

The map of marginal excursion probabilities for the level $50 \mu g/m^3$ for January 30, 2006, based on the estimated posterior distribution for $x$, can be seen in the left panel of Figure \ref{fig:pm10}. To avoid inappropriate linear extrapolation of the effect of elevation beyond the range of the elevation of the observations, the results are only shown for areas below 1km. Based on these results, we now estimate the positive excursion function for the level $50 \mu g/m^3$, $F_{50}^+(\mv{s})$, using the NI method from Section \ref{sec:inlaprob} and the parametric family of excursion sets from Definition \ref{def:parametric0}. A total of $25$ parameter configurations are used in the integration. The result can be seen in the right panel of Figure \ref{fig:pm10}. As seen in the figure, there are three regions where the level is clearly exceeded, and a fourth that possibly contains too high pollution levels. As expected, these areas coincide with the locations of the main metropolitan areas in the region; Turin, Novara, Vercelli, and Alessandria. In this case, it would have been desirable to make the predictions on a finer spatial scale, but since the covariates were given on a $4$ km $\times$ $4$ km grid, this spatial resolution had to be used in the prediction.

\begin{figure}
\begin{center}
\begin{minipage}[b]{0.35\linewidth}
\centering
Marginal probabilities\\
\includegraphics[width=\linewidth]{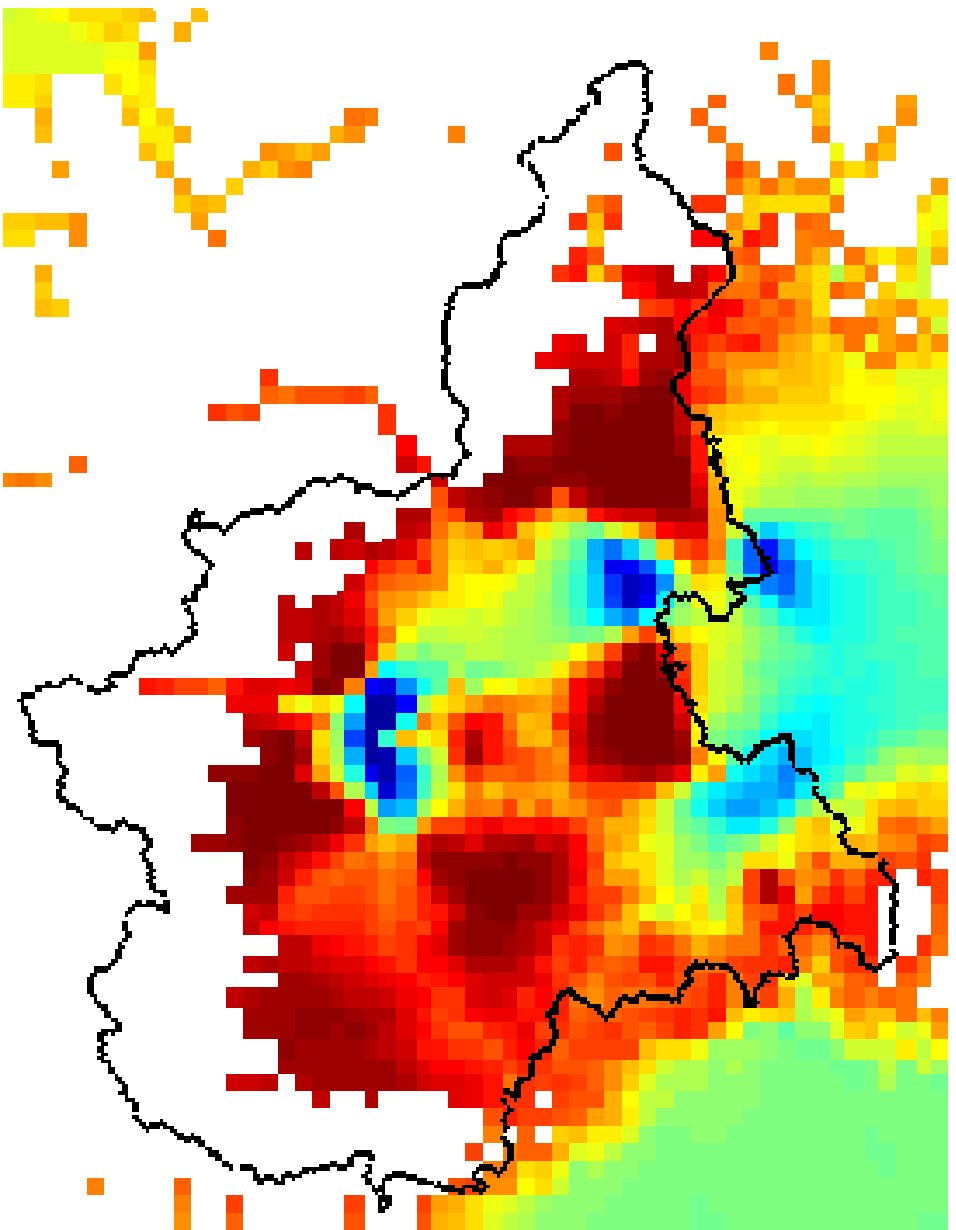}%
\end{minipage}
\begin{minipage}[b]{0.35\linewidth}
\centering
\phantom{p}Excursion function $F_{50}^-(\mv{s})$\phantom{p}\\
\includegraphics[width=\linewidth]{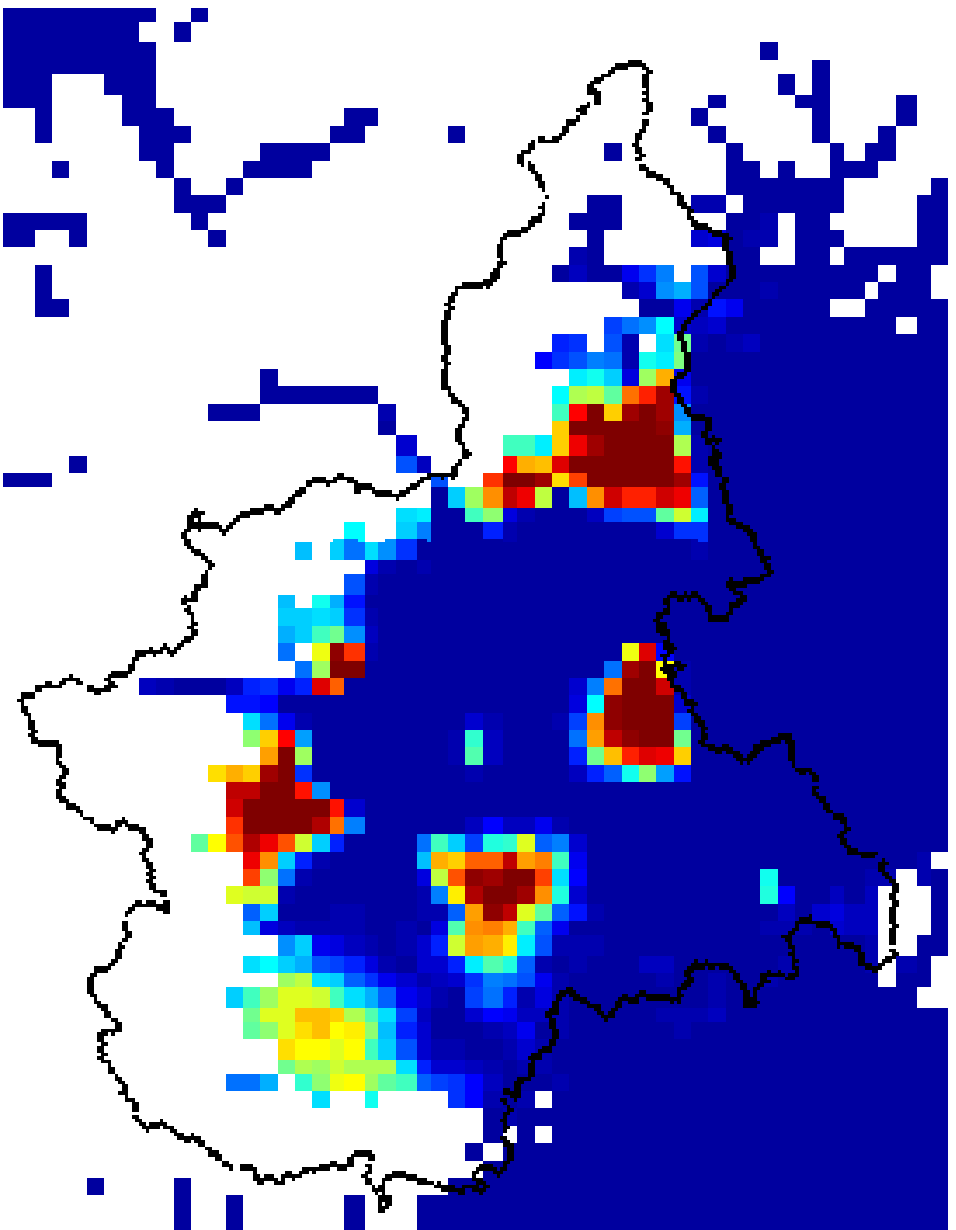}%
\end{minipage}
\begin{minipage}[b]{0.05\linewidth}
\centering
\quad\\
\input{figs/pm10_cbar.tex}
\includegraphics[height=81mm,bb= 65 2860 90 3063,clip=]{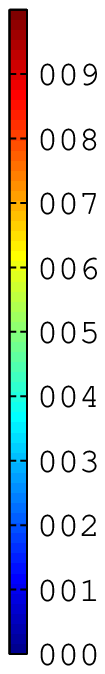}%
\end{minipage}
\end{center}
\vspace{-0.4cm}
\caption{Results from the PM$_{10}$ application for January 30, 2006. A map of the marginal probabilities for the field being below the level $50 \mu g/m^3$ (left), and the joint negative excursion distribution function for the level (right).}
\label{fig:pm10_2}
\end{figure}

To get a better understanding of the results, it is also of interest to find the regions where the pollution level is simultaneously below the limit value with some given probability. The marginal probabilities for being below be the level $50 \mu g/m^3$, based on the estimated posterior distribution for $x$, can be seen in the left panel of Figure \ref{fig:pm10_2}. The results are again only shown for areas below 1km altitude. Using the same method as for the positive excursion function, we now estimate the negative excursion function for the level $50 \mu g/m^3$, $F_{50}^-(\mv{s})$. The result can be seen in the right panel of Figure \ref{fig:pm10_2}. 

\begin{figure}[t]
\begin{center}
\begin{minipage}[b]{0.35\linewidth}
\centering
\quad\quad\quad Excursion sets\phantom{$F_{50}^c(\mv{s})$}\\
\includegraphics[width=\linewidth]{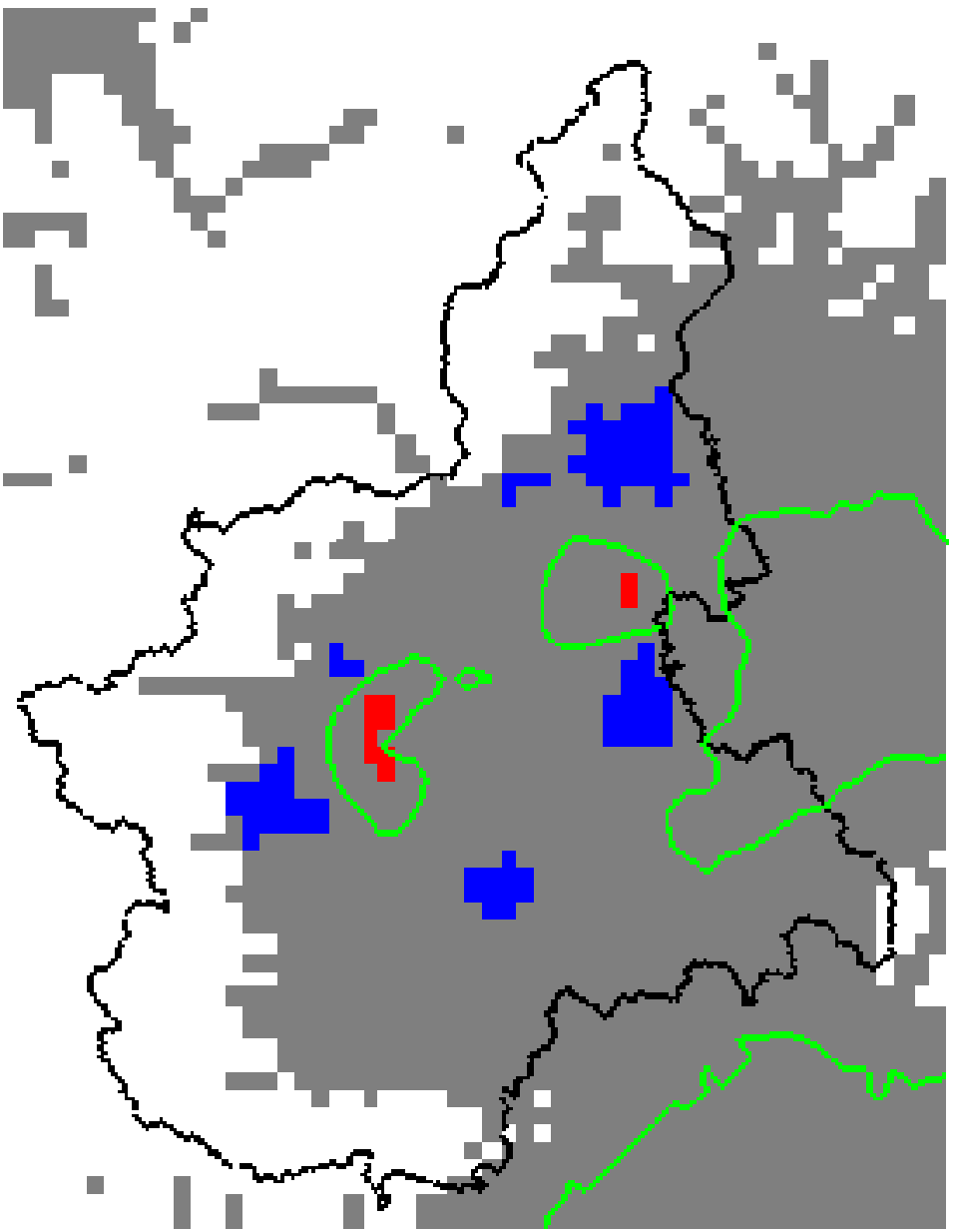}%
\end{minipage}
\begin{minipage}[b]{0.35\linewidth}
\centering
Contour function $F_{50}^c(\mv{s})$\\
\includegraphics[width=\linewidth]{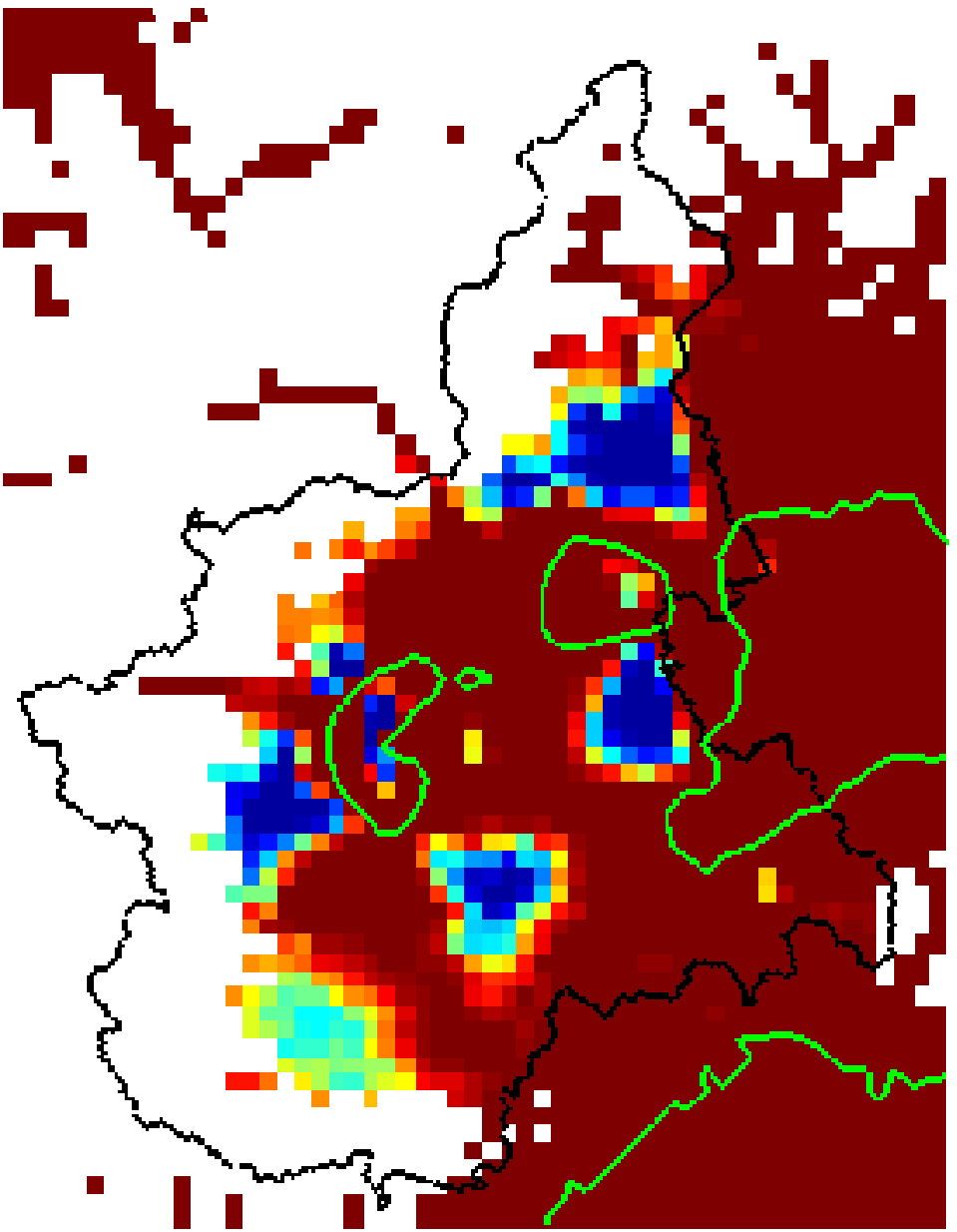}%
\end{minipage}
\begin{minipage}[b]{0.05\linewidth}
\centering
\quad\\
\input{figs/pm10_cbar.tex}
\includegraphics[height=81mm,bb= 65 2860 90 3063,clip=]{figs/pm10_cbar2.eps}%
\end{minipage}
\end{center}
\vspace{-0.4cm}
\caption{Results from the PM$_{10}$ application for January 30, 2006. In the left panel, the set $\exset{50,0.1}{+}(x)$ is shown in red, $\exset{50,0.1}{-}(x)$ in blue, and it's complement $\exset{50,0.1}{+}(x)^{c}$ in grey. The contour curve for the level $50 \mu g/m^3$ is shown in green. The right panel shows the contour function for the level $50 \mu g/m^3$, $F_{50}^c(s)$.}
\label{fig:pm10_3}
\end{figure}

Note that the union of $\exset{50,0.1}{+}(x)$ and $\exset{50,0.1}{-}(x)$ covers only a small part of the region, indicating that the uncertainty in the problem is large. See the red and blue sets in the left panel of Figure \ref{fig:pm10_3}. Also, by taking the complement of the set $\exset{50,0.1}{-}(x)$, we get the region that contains all exceedances of the level with certainty $0.9$, indicated in grey in the left panel of Figure \ref{fig:pm10_3}. This set is large, indicating that there are many regions where the level possibly is exceeded. Hence, it is important to note that the positive excursion set $\exset{50,0.1}{+}(x)$ is small because the uncertainty is large in the problem, and not because the other regions certainly have concentrations below the level. 

To verify that the uncertainty is large, we finally calculate the contour function for the level $50 \mu g/m^3$, $F_{50}^c(\mv{s})$, using the NI method and the one-parameter family from Definition \ref{def:paravoiding} for the pair of level avoiding sets. The result can be seen in the right panel of Figure \ref{fig:pm10_3}, and the uncertainty region for the contour curve indeed covers a large part of the region which indicates that the uncertainty in the estimated contour curve is large. 

\subsection{Spatially dependent temporal trends in vegetation data}\label{paperE:sec:sahel}
Trends in vegetation cover are related to changes in climatic drivers, feedback mechanisms between the atmosphere and land surface, and human interaction. A region with rapid recent changes is the African Sahel. This zone has received much attention regarding desertification and climatic variations \citep{olsson, nicholson00, lamb}. Recently, \citet{eklundh} observed a strong increase in seasonal vegetation index over parts of the Sahel using Advanced Very High Resolution Radiometer (AVHRR) data from the NOAA/NASA Pathfinder AVHRR Land (PAL) database \citep{NOAA/NASA, james}, for the period 1982-1999. The study 
was based on ordinary least squares linear regression on individual time 
series extracted for each pixel in the satellite images. The results of \citet{eklundh} were later improved by \cite{bolin09a} where a spatial model for the vegetation was used in the analysis to capture the spatial dependencies in the trend estimation.

To find regions where changes in the vegetation have occurred over the course of the studied time period, both \citet{eklundh} and \cite{bolin09a} used significance testing for the individual pixels in the field. Thus, pixels that individually had significant changes in vegetation were found, but no attempts were made to find simultaneous excursion regions. Here, we will use a similar model to that of \cite{bolin09a} but also estimate joint excursion regions for the vegetation trends. 

Assume that the vegetation measurements year $t$ are generated as,
\begin{equation*}
\mv{Y}_t | \mv{X}_t,\mv{\Sigma}_{\ep_t} \in \pN(\mv{A}_t\mv{X}_t,
\mv{\Sigma}_{\ep_t}),
\end{equation*}
where $\mv{X}_t$ is the latent vegetation field with
prior distribution $\pi(\mv{X}_t)$, $\mv{\Sigma}_{\ep_t}$ is a
measurement noise covariance matrix, and $\mv{A}_t$ is an
observation matrix determining which pixels in the field that are observed. To estimate time varying trends in the observations, $\mv{X}$ is restricted to follow a field of spatially varying linear trends:
\begin{equation}\label{B}
\mv{X}_t = \mv{K}_1 + t\mv{K}_2
\end{equation}
The prior distribution for $\mv{K} = [\mv{K}_1^\trsp, \mv{K}_2^\trsp]^\trsp$ is obtained by evaluating the joint distribution for $\mv{X} = [\mv{X}_1^\trsp,\ldots, \mv{X}_T^\trsp]^\trsp$ conditionally on the restriction \eqref{B}. We choose a second-order polynomial IGMRF \citep[][Section
3.4.2]{rue1} prior for $\mv{X}$ and then calculate the corresponding prior distribution for $\mv{K}$, \citep[see][for details]{bolin09a}.

To complete the model, the
structure of $\mv{\Sigma}_{\ep}$ needs to be determined. Many of the factors that the
measurement noise should model are local phenomena, such as aerosol
and cloud cover. Since it seems unreasonable that the scale of these
disturbances would be the same over the entire region, \cite{bolin09a} assumed that the measurement noise was uncorrelated with a different noise variance at each pixel in the field. This results in a large number of parameters for the measurements noise, one for each pixel in the field, so here we instead use a different slightly simplified noise model. We divide the region into five different land cover categories using the Africa Land Cover Characteristics Data Base Version 2.0 (\texttt{http://edc2.usgs.gov/glcc/glcc.php}):
\begin{inparaenum}[\itshape 1\upshape)]
\item Bare desert;
\item Semi desert;
\item Savanna;
\item Crops, grass, and shrubs; and
\item Forests and wetlands.
\end{inparaenum}
The measurement noise variance at pixel $\mv{s}_i$ is then modeled as
\begin{equation}\label{eq:sigma}
\log \sigma^2(\mv{s}_i) = \sum_{k=1}^5 \theta_k b_k(\mv{s}_i),
\end{equation}
where $b_k(\mv{s})$ is the spatial basis function with values equal to the proportion of vegetation type $k$ at each pixel $\mv{s}$. The parameters of the model are thus the scale parameter $\kappa$ and the five measurement noise parameters $\theta_1,\ldots, \theta_5$. A gamma prior is assumed for $\kappa$ and gaussian priors are used for $\theta_k$.

We choose to study the western part of the Sahel region, and this area is divided into $35463$ pixels of size $8$ km $\times$ $8$ km, so the field $\mv{K}$ has $70926$ elements, and there are $547832$ measurements from $17$ years of data starting in 1982 and ending in 1999.

\begin{figure}
\begin{center}
\begin{minipage}[b]{0.9\linewidth}
\centering
$\mv{K}_1$\\
\includegraphics[width=\linewidth,bb=92 343 543 468,clip=]{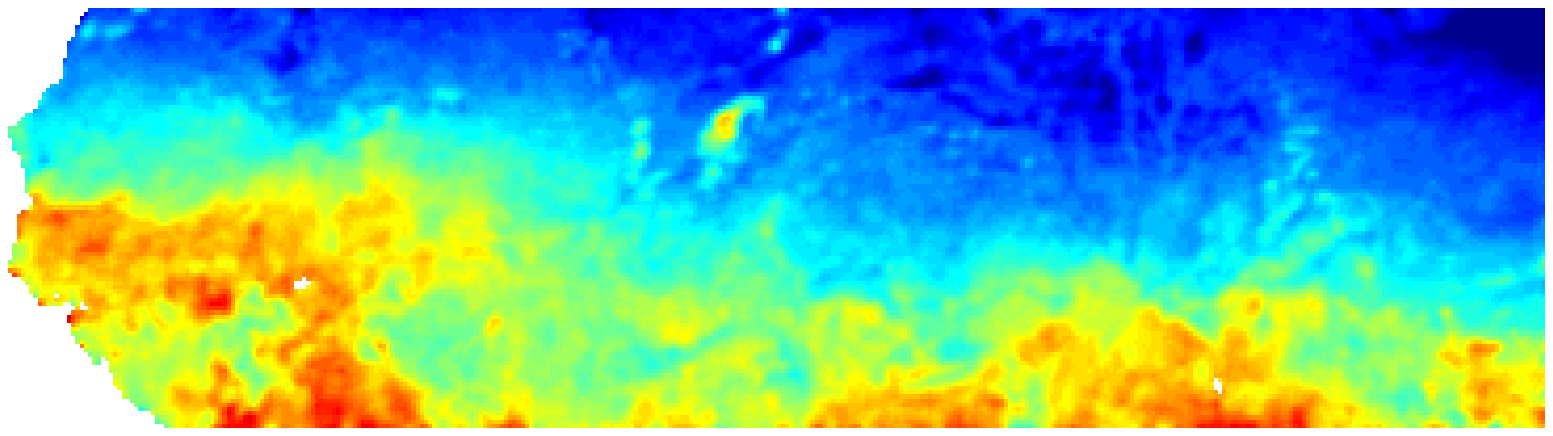}%
\end{minipage}
\begin{minipage}[b]{0.05\linewidth}
\centering
\quad\\
\input{figs/sahel_k1cv.tex}
\includegraphics[height=44.5mm, bb=65 2848 102 2949,clip=]{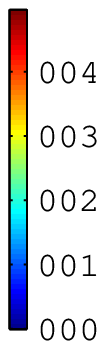}%
\end{minipage}\\
\vspace{0.1cm}
\begin{minipage}[b]{0.9\linewidth}
\centering
$\mv{K}_2$\\
\includegraphics[width=\linewidth,bb=92 343 543 468,clip=]{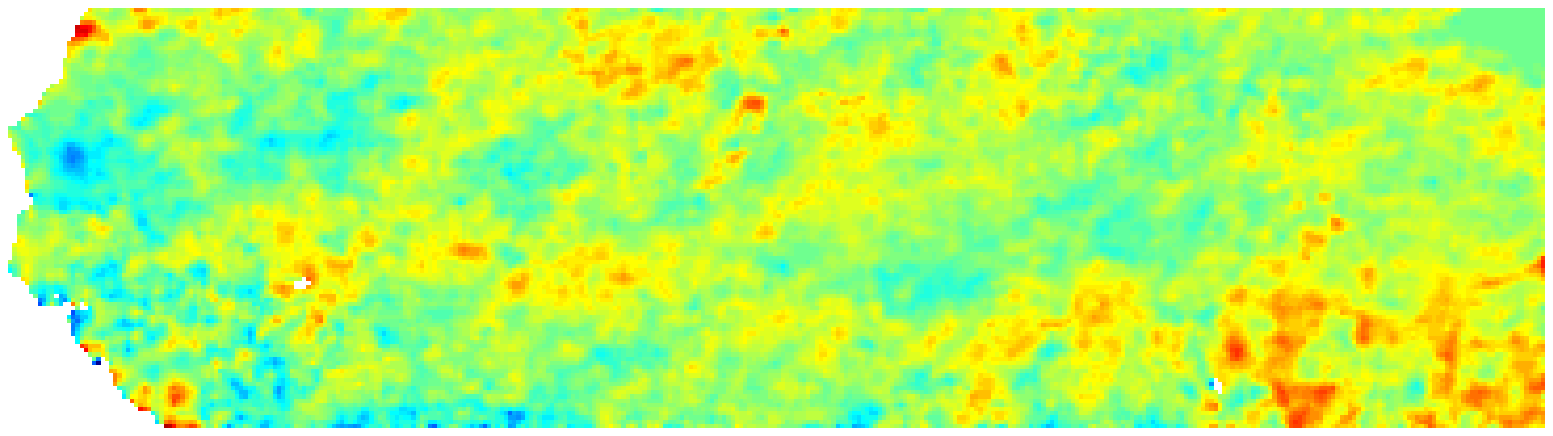}%
\end{minipage}
\begin{minipage}[b]{0.05\linewidth}
\centering
\quad\\
\input{figs/sahel_k2cv.tex}
\includegraphics[height=44.5mm, bb=65 2848 102 2949,clip=]{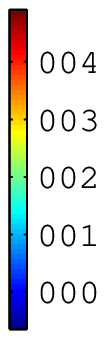}%
\end{minipage}
\end{center}
\vspace{-0.6cm}
\caption{Results from the Sahel vegetation data. The top panel shows the posterior estimates of the regression intercepts, $\mv{K}_1$, and the bottom panel shows the estimated slopes, $\mv{K}_2$.}
\label{fig:sahel}
\end{figure}

\begin{figure}
\begin{center}
\begin{minipage}[b]{0.9\linewidth}
\centering
$\sigma(\mv{s})$\\
\includegraphics[width=\linewidth,bb=92 343 543 468,clip=]{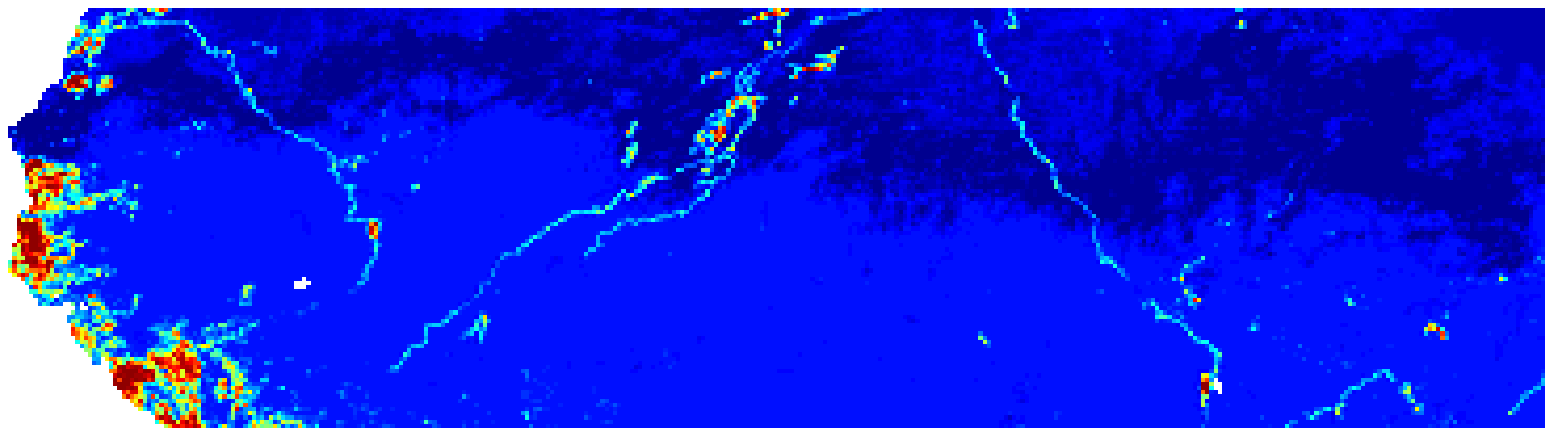}%
\end{minipage}
\begin{minipage}[b]{0.05\linewidth}
\centering
\quad\\
\input{figs/sahel_scv.tex}
\includegraphics[height=44.5mm, bb=65 2848 102 2949,clip=]{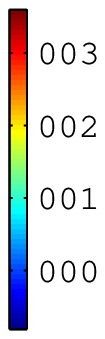}%
\end{minipage}\\
\vspace{0.1cm}
\begin{minipage}[b]{0.9\linewidth}
\centering
Excursion set\\
\includegraphics[width=\linewidth,bb=92 286 543 468,clip=]{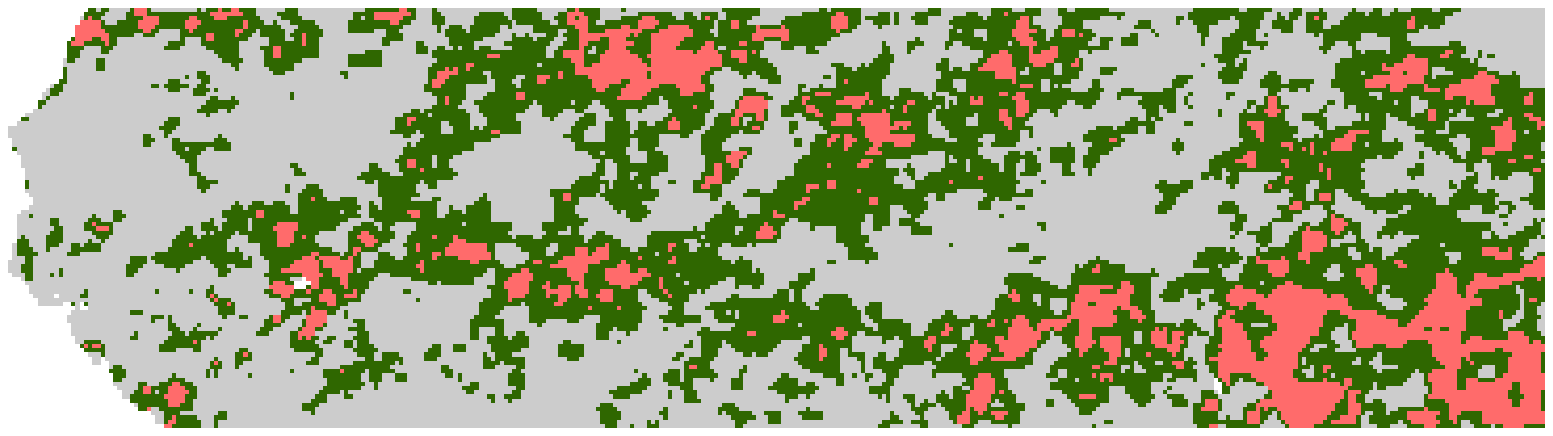}%
\end{minipage}
\begin{minipage}[b]{0.05\linewidth}
\centering
\quad\\
\includegraphics[height=44.5mm, bb=102 2848 102 2949,clip=]{figs/sahel_scv.eps}%
\end{minipage}
\end{center}
\vspace{-2cm}
\caption{Results from the Sahel vegetation data. The top panel shows the estimated standard deviation of the measurement noise and the bottom panel shows the estimated excursion set $\exset{0,0.05}{+}(\mv{K}_2)$ in red and the point-wise positive significant trends in green.}
\label{fig:sahel2}
\end{figure}

The model parameters and the marginal posterior distributions are estimated using the INLA framework and the excursion sets are estimated using the QC~method from Section \ref{sec:inlaprob}. The results can be seen in Figure \ref{fig:sahel} and Figure~\ref{fig:sahel2}. The top panel in Figure \ref{fig:sahel} shows the posterior estimates of the intercepts, $\mv{K}_1$, and the slopes, $\mv{K}_2$, is shown in the bottom panel. As expected, intercepts are larger in the savanna regions to the south, and smaller in the semi desert regions to the north. The top panel of Figure \ref{fig:sahel2} shows the estimated standard deviation of the measurement noise using model \eqref{eq:sigma}. It is worth noting that these results look reasonable, with larger measurement errors in the coastal region and where there are forests and wetlands, and smaller measurement errors in desert and semi desert regions. Finally the bottom panel of Figure \ref{fig:sahel2} shows the estimated excursion set $\exset{0,0.05}{+}(\mv{K}_2)$ in red and the point-wise positive significant trends in green. The interpretation of the result is that one with high certainty can conclude that the areas indicated in red have experienced an increase in vegetation over the studied time period. Hence, conclusions drawn by \cite{eklundh} seem valid, also when taking the spatial dependency of the vegetation into account and when estimating the excursion sets controlling the family-wise error.

\section{Discussion}\label{sec:conclusions}
Estimating excursion sets and uncertainty regions for contour curves for stochastic fields are difficult problems, both because of computational issues but also because it might not be clear how such uncertainty regions should be defined. In this work, we have given precise definitions for these uncertainty regions, introduced the concept of excursion functions as a visual tool for illustrating the uncertainty in these regions, and presented a method for calculating these quantities for latent Gaussian models. 

The main idea behind the computational method is to use a parametric family for the excursion sets in combination with a sequential integration method to reduce the computational effort required when estimating the sets in practice. Tests on simulated data showed that the method is accurate, and two applications were presented to show that the method is applicable even to large environmental problems. 

There are a number of extensions that could be made to this work. First of all, using the one-parameter family for the excursion sets gives a method that falls into the broad category of $p$-value thresholding methods for estimating simultaneous excursion sets. As previously mentioned, the important advantage with the method proposed here compared with other commonly used thresholding methods is that the correct joint distribution is used when selecting the threshold. The disadvantage is that the method is computationally more expensive than many of the standard thresholding methods. It would, therefore, be interesting to do a comparison with other similar methods with respect to the accuracy and computational complexity. Another interesting comparison would be to compare the uncertainty regions for contour curves produced by these methods to those of \cite{Lindgren95}. One could potentially also combine these methods with the work by \cite{Polfeldt99} to make statements on the quality of contour maps.   

We also presented other parametric families that can be used to obtain more complicated methods for estimating the excursion sets, with the possibility of finding more precise estimates under the cost of higher computational complexity. Initial comparisons showed that there is not much gain in using these more complicated methods, but so far these comparisons have only been made using fairly simple latent models, and the gain is likely higher when the latent models are more complex. Hence, more studies are required to verify if this is the case and to investigate in what situations it is appropriate to use the simple one-parameter families. One possible advantage with the more complicated parametric families is when one has prior knowledge regarding the shape of the excursion sets. For example, if one knows that the excursion sets should be large contiguous regions, such knowledge could be incorporated using a two-parameter smoothing family.

As a final note, the method introduced here is available as a C-package with interfaces to both R and Matlab, see the supplementary material for the online version of the article for details. 

\section*{Acknowledgements}
Data used by the authors in the Sahel vegetation study include data produced through funding from the Earth Observing System Pathfinder Program of NASA's
Mission to Planet Earth in cooperation with National Oceanic and
Atmospheric Administration. The data were provided by the Earth
Observing System Data and Information System (EOSDIS), Distributed
Active Archive Center at Goddard Space Flight Center which archives,
manages, and distributes this data set. The data used in the PM$_{10}$ study was provided by the information system Aria Web Regione Piemonte and Arpa Piemonte. The authors are grateful to Johan Lindström and Daniel Simpson for valuable discussions on the subject of excursions and contour curve uncertainty sets, and to Peter Guttorp for highlighting the need for a thorough treatment of the subject.

\begin{appendix}
\section{Notes on the MCMC algorithm used in Example 3}\label{sec:appendixa}
To generate samples from the posterior distribution $\pi(\mv{x}|\mv{y})$ in Example 3, an MCMC algorithm is used. The algorithm is a random-walk Metropolis Hastings algorithm \citep{Metropolis53,Hastings70} with proposal kernel
\begin{equation*}
q(\{\mv{x}_{\mbox{\footnotesize old}},\mv{\theta}_{\mbox{\footnotesize old}}\},\{\mv{x},\mv{\theta}\}) = \pi(\mv{x}|\mv{y},\mv{\theta})q_{\theta}(\mv{\theta}_{\mbox{\footnotesize old}},\theta).
\end{equation*}
A new proposal of the parameters $\mv{\theta} = (\log(\sigma), \log(\kappa), \log(\phi))$ is proposed based on the old value $\mv{\theta}_{\mbox{\footnotesize old}}$ using $\mv{\theta} \sim \pN(\mv{\theta}_{\mbox{\footnotesize old}}, \mv{\Sigma}_{\theta})$. Here $\mv{\Sigma}_{\theta}$ is a scaled version of the Hessian matrix evaluated at the maximum posterior estimate of $\mv{\theta}$. The scaling is selected as suggested by \cite{gelman96}. A new value for $\mv{x}$ is then proposed using the marginal posterior distribution $\pi(\mv{x}|\mv{y},\mv{\theta})$ given by $\mv{x}|\mv{y},\mv{\theta} \sim \pN(\frac{1}{\sigma^2}\hat{\mv{Q}}^{-1}\mv{A}^{\trsp}\mv{y},\hat{\mv{Q}})$. Here $\hat{\mv{Q}} = \mv{Q} +  \frac{1}{\sigma^2}\mv{A}^{\trsp} \mv{A}$, where $\mv{Q}$ is the precision matrix for $\mv{x}$ and $\mv{A}$ is an observation matrix determined by the measurement locations. The acceptance probability simplifies to
\begin{equation*}
\alpha_{\mbox{\footnotesize MCMC}} = \min\left(1,\frac{\pi(\mv{\theta}_{\mbox{\footnotesize old}}|\mv{y})}{\pi(\mv{\theta}|\mv{y})}\right),
\end{equation*}
where the posterior $\pi(\mv{\theta}|\mv{y})$ is given by
\begin{equation}
\pi(\mv{\theta}|\mv{y}) \propto
  \frac{|\mv{Q}|^{\frac{1}{2}}\pi(\mv{\theta})}{|\hat{\mv{Q}}|^{\frac{1}{2}} |\sigma \mv{I}|} \exp\left(\frac{1}{2\sigma^2}\mv{y}^{\trsp}\left(\frac{ \mv{A}\hat{\mv{Q}}^{-1} \mv{A}^{\trsp}}{\sigma^2} - \mv{I}\right)\mv{y} \right).
\end{equation}
Since the proposal for $\mv{x}$ does not affect the acceptance probability, a new proposal for $\mv{x}$ is only generated if $\mv{\theta}$ is accepted. With only three parameters in the model, we achieve good mixing this way, but being an MCMC-procedure it is still highly computationally demanding since the calculation of the acceptance probability requires a few Cholesky factorizations and back substitutions based on the posterior precision matrix for $\mv{x}$. 
\end{appendix}

\bibliographystyle{plainnat}
\bibliography{journ_abrv,completebib}
\end{document}